\documentclass[aps,pra,superscriptaddress,floatfix,amsmath]{revtex4-2} 
\usepackage{multirow,dsfont,color}
\usepackage{graphicx}
\usepackage{hyperref}
\usepackage{braket}
\usepackage{natbib}
\usepackage{setspace}
\usepackage{subcaption}

\def\equationautorefname~#1\null{(#1)\null}

\newcommand{\unit}{\mathds{1}}

\newcommand{\G}{\mathbf{G}}

\begin{document}

\title{Improving Hamiltonian encodings with the Gray code}
\author{Olivia Di Matteo}
\author{Anna McCoy} 
\affiliation{TRIUMF, Vancouver, British Columbia V6T 2A3, Canada}

\author{Peter Gysbers}
\affiliation{TRIUMF, Vancouver, British Columbia V6T 2A3, Canada}
\affiliation{Department of Physics and Astronomy, University of British Columbia, Vancouver, British Columbia V6T 1Z1, Canada}

\author{Takayuki Miyagi} 
\author{R. M. Woloshyn}
\author{Petr Navr\'{a}til} 
\affiliation{TRIUMF, Vancouver, British Columbia V6T 2A3, Canada}

\date{\today}

\begin{abstract}

 Due to the limitations of present-day quantum hardware, it is especially critical to design algorithms that make the best possible use of available resources. When simulating quantum many-body systems on a quantum computer, straightforward encodings that transform many-body Hamiltonians into qubit Hamiltonians use $N$ of the available basis states of an $N$-qubit system, whereas $2^N$ are in theory available. We explore an efficient encoding that uses the entire set of basis states, where terms in the Hamiltonian are mapped to qubit operators with a Hamiltonian that acts on the basis states in Gray code order. This encoding is applied to the commonly studied problem of finding the ground-state energy of a deuteron with a simulated variational quantum eigensolver (VQE). It is compared to a standard ``one-hot" encoding, and various trade-offs that arise are analyzed. The energy distribution of VQE solutions has smaller variance than the one obtained by the one-hot encoding even in the presence of simulated hardware noise, despite an increase in the number of measurements. The reduced number of qubits and a shorter-depth variational ansatz enables the encoding of larger problems on current-generation machines. This encoding also demonstrates improvements for simulating time evolution of the same system, producing circuits for the evolution operators with reduced depth and roughly half the number of gates compared to a one-hot encoding.
\end{abstract}
\maketitle

\section{Introduction}

The simulation of quantum many-body systems remains a complex and computationally challenging problem in physics and quantum chemistry.  Direct solutions are often limited by the rapid growth of the problem as the number of particles and relevant degrees of freedom increases. Quantum computers may play a key role in overcoming these computational challenges ~\cite{Feynman1982, Lloyd1996}; however it is only with recent advances in qubit technology~\cite{Kjaergaard2020,Bruzewicz2019,Bromley2020} that such a use of quantum computers has begun to be feasible~\cite{Lanyon2010,Peruzzo2014,OMalley2016,Kandala2017,Dumitrescu2018,Klco2018,Shehab2019,McCaskey2019,Nam2020}.

Quantum many-body problems which are typically solved in a configuration interaction (CI) framework~\cite{Lowdin1999,Mayer1949-75,Mayer1949-78,Barrett2013,Maris2009} may be well suited for quantum computers.  In a CI framework, the wavefunction is expanded in terms of an occupation basis, where each basis state corresponds to a distribution of the particles over the different possible single-particle substates.  In nuclear physics applications, these substates are typically taken to be harmonic oscillator orbitals. 

As the number of particles and/or the number of included orbitals increases, the number of basis states $N$ grows exponentially. This can limit the size of the systems that can be studied. Since on a quantum computer $N$ states can be mapped to as few as $\lceil \log_2 N \rceil$ qubits this may, in the future, allow for many-body calculations on a scale larger than is feasible on classical computers. The challenge, then, is to map the physical basis states and operators onto qubits and quantum circuits in an efficient manner.  

The methods available to map the many-body problem to a quantum computer depend on how the Hamiltonian is expressed \cite{mcardle2018quantum,Jordan1928, Somma2002, Somma2003,Bravyi2002, Seeley2012, Tranter2015,Sawaya2020}. In this work, the simplest nuclear many-body problem, that of the deuteron (consisting of a neutron and a proton), is considered. Following \cite{Dumitrescu2018,Shehab2019}, the Hamiltonian describing the relative motion of the neutron and proton is expressed in terms of matrix elements in a harmonic oscillator basis. In
\cite{Dumitrescu2018,Shehab2019} a one-hot mapping of harmonic oscillator basis states to qubits was used, requiring a number of qubits equal to the number of basis states.

At the other end of the encoding spectrum, binary encodings allow many-body bases with $N$ states to be represented in terms of $\lceil \log_2 N \rceil$ qubits. While many different binary encodings are possible \cite{Bravyi2017, McArdle2019, Sawaya2019, Sawaya2020, Kyaw2020, Kottmann2020}, this work emphasizes one based on a \emph{Gray code} \cite{gray1953patent}, and was inspired by an earlier investigation of quantum simulation of a lattice gauge theory \cite{Lewis2019}.  Application of Gray codes in Hamiltonian encodings and Hamiltonian simulation was recently explored in the work of \cite{Sawaya2020}, where it is noted that Gray code encodings are particularly resource-efficient for tridiagonal Hamiltonians which is the case for the deuteron problem addressed here. 

This work performs an analysis of trade-offs that arise between one-hot and Gray code encodings for the deuteron problem, starting from the level of Pauli terms in the Hamiltonian and down to simulation of noisy hardware devices. While the Hamiltonians constructed with the Gray code encoding have more Pauli terms (and require more measurement settings), the reduction in both the number of qubits and the number of controlled-NOT (CNOT) gates in circuits for various applications has important consequences in the noisy intermediate-scale quantum (NISQ) computing era~\cite{Preskill2018}. 
 The low coherence times and high gate error rates of current quantum computers make direct solution of the deuteron eigenproblem on a quantum computer infeasible. However, a number of hybrid quantum-classical algorithms have emerged over the past decade, where an optimization problem run on a classical computer is assisted by a quantum computer that can compute its cost function more efficiently. One such algorithm is the variational quantum eigensolver (VQE)~\cite{Peruzzo2014, Wecker2015, McClean2016}, which can be used to find the ground-state energy of a Hamiltonian~\cite{Peruzzo2014}. Here a simulated VQE is used to obtain the ground-state energy of the deuteron. Results are obtained for both the Gray code encoding and the one-hot encoding and compared. As we will show, the variance in distribution of energies found from the VQE is significantly smaller for the Gray code encoding, most notably in cases with simulated hardware noise.

Looking beyond the NISQ era, the problem of finding the ground-state energy can also be addressed using quantum phase estimation. Phase estimation circuits have depth beyond what is feasible for a NISQ machine, and require the implementation of a unitary which simulates the time evolution of the system, often termed Hamiltonian simulation~\cite{Lloyd1996}. This work presents an end-to-end, hardware-aware analysis of the Gray code and one-hot encodings applied to Hamiltonian simulation of the deuteron system. It finds reduced circuit depth and gate count of the unitary evolution operators of the Gray code encoding compared to those obtained using the one-hot encoding, and through noisy simulations it demonstrates the potential for larger, more accurate problems to be run on NISQ-era devices. 

\autoref{sec:many-body-problems} presents the deuteron Hamiltonian and the resultant qubit Hamiltonian under the one-hot encoding. The Gray code encoding is introduced and applied to the deuteron in \autoref{sec:gray-code}, followed by analysis of its structure in a more general setting. \autoref{sec:implementation} details the implementation of the variational quantum eigensolver and simulation results of computational experiments with shot noise, and simulated hardware noise. \autoref{sec:hamsim} presents the implementation of the Gray code encoding for Hamiltonian simulation. Conclusions and future directions are discussed in \autoref{sec:conclusions}.

\section{Encoding quantum many-body problems on a quantum computer}
\label{sec:many-body-problems}

\subsection{A Hamiltonian for the deuteron}
\label{subsec:deuteron}

Atomic nuclei are self-bound systems with interactions among nucleons, the building blocks of the nucleus, depending on nucleon relative positions and momenta, as well as their spins and isospins. The corresponding nuclear Hamiltonian is then translationally invariant and the relative or Jacobi coordinates and momenta form a natural coordinate system to use. In the special case of the deuteron, there is a single relative coordinate $\vec{r}{=}\vec{r}_1{-}\vec{r}_2$ and the canonical relative momentum $\vec{p}{=}\frac{1}{2}(\vec{p}_1{-}\vec{p}_2)$. To model the proton-neutron interaction, we follow Refs.~\cite{prc-93-2016-044332-Binder,prc-98-2018-054301-Bansal} and apply the pionless effective field theory (EFT). To leading order in pionless EFT, the constituent proton and neutron interact via a short-ranged contact interaction in the $^3S_1$ partial wave ($L{=}0$ and $S{=}1$, $J{=}1$). Using a harmonic oscillator (HO) basis expansion of the trial wave function that depends on $\vec{r}$, only the radial part $R_{n L{=}0}(r,\omega)$ remains relevant, with $\omega$ the HO frequency. In this basis, the deuteron Hamiltonian, $H{=}T{+}V$ with $T$ the kinetic and $V$ the potential energy, is defined by
\begin{equation}
    H_N = \sum^{N-1}_{n,n'=0} \braket{n'|(T + V)|n} \ket{n'} \bra{n},
    \label{eq:basic_ham}
\end{equation}
where
\begin{subequations}
\begin{equation}
\braket{n' | T |n} = \frac{\hbar \omega}{2} \left[ (2n + 3/2) \delta^{n'}_n - \sqrt{n(n+1/2)}\delta^{n'}_{n-1} - \sqrt{(n+1)(n+3/2)} \delta^{n'}_{n+1}  \right]  ,
 \end{equation}
 \begin{equation}
\braket{ n' | V |n} = V_0 \delta^0_n \delta^{n'}_n \;.
\end{equation}
\label{eq:matel}
\end{subequations}

The $n=0,1,\dots, N-1$ is the \emph{relative} harmonic oscillator radial node number. 
For the calculations used in this paper, the harmonic oscillator basis parameter is chosen to be $\hbar \omega = 7$ MeV and thus $V_0 = -5.68658111$ MeV following Ref~\cite{Dumitrescu2018}. 

The Hamiltonian is truncated to include only states with $n<N$.  As the size of the harmonic oscillator basis increases with $N$, the eigenvalues of each Hamiltonian $H_N$ converge towards the true ground-state energy.  With the selected model parameters~\cite{Dumitrescu2018}, in the limit $N \rightarrow \infty$, the ground-state energy of the deuteron fits its experimental value: $-2.224 $ MeV.


\subsection{Mapping using a one-hot encoding}
\label{subsec:jordan-wigner}

To solve a Hamiltonian eigenproblem using a quantum computer, the many-body basis and relevant operators must be reexpressed in the language of qubits. The first step is to make a mapping from the original basis states (in the harmonic oscillator basis) to qubit basis states. This can be done in any order, though a straightforward choice is a ``one-hot," or unary, encoding,
\begin{equation}
 \begin{aligned}
    \ket{0} &\rightarrow \ket{1000},\quad&
    \ket{1} &\rightarrow \ket{0100},\quad&
    \ket{2} &\rightarrow \ket{0010},\quad&
    \ket{3} &\rightarrow \ket{0001}.
 \end{aligned}
\end{equation}
Under this encoding, the $ith$ state is mapped to the $ith$ qubit.  In the deuteron example, this corresponds to mapping the $N$ \emph{relative} states to $N$ qubits. 

After choosing a mapping, we must rewrite the Hamiltonian over the $N$-qubit Pauli group $\mathcal{P}_N$. The new Hamiltonian must have the analogous action on the qubit basis states as the original Hamiltonian of Eq.~\autoref{eq:basic_ham} has on the harmonic oscillator basis states. As the original Hamiltonian is tridiagonal, this requires constructing components that act as number operators (to replace the $|n\rangle \langle n|$ terms), as well as ladder operators (to replace the $|n+1 \rangle \langle n|$ terms). 

Recall that $\mathcal{P}_N$ is generated by $N$-fold tensor products of $Z$ and $X$, where
\begin{eqnarray}
    Z =  \begin{pmatrix}
     1 & 0 \\ 0 & -1
    \end{pmatrix}, &\quad& Z \ket{0} = \ket{0}, \enskip Z \ket{1} = -\ket{1},\\ 
    X =  \begin{pmatrix}
     0 & 1 \\ 1 & 0
    \end{pmatrix}, &\quad& X \ket{0} = \ket{1}, \enskip X \ket{1} = \ket{0}.
\end{eqnarray}
We use the notation $\{X_i, Y_i, Z_i\}$, where $X_i$, $Y_i$, and $Z_i$ represent the application of the associated Pauli on qubit $i$, and identity on all unspecified qubits. 

To construct the number-type components of the Hamiltonian, we employ projection operators $P^{(m)}\ket{n}=\delta_{mn}\ket{n}$, for $m,n=0,1$. These can be expressed in terms of the $Z$ operator as 
\begin{equation}
    P^{(0)} = \frac{1}{2} \left( \unit + Z \right) = \begin{pmatrix}
     1 & 0 \\ 0 & 0
    \end{pmatrix}, \qquad
       P^{(1)} =  \frac{1}{2} \left( \unit - Z \right) =\begin{pmatrix}
     0 & 0 \\ 0 & 1
    \end{pmatrix}.
\end{equation}
Note that these correspond to $|0\rangle \langle 0|$ and $|1\rangle\langle1|$, respectively. Each number-type operator is thus obtained from a term with $P^{(1)}$ on qubit $n$ prefixed with the appropriate coefficient $\langle n | H | n \rangle$ obtained from Eq.~\autoref{eq:matel}.

The ladder-type terms require operators that correspond to transforming the states of two adjacent qubits between $|10\rangle \leftrightarrow |01\rangle$. At the individual qubit level, this applies  $|0\rangle \langle 1|$ to the first qubit and $|1\rangle \langle 0|$ to the adjacent qubit. These projectors are equal to $\frac{1}{2}(X + iY)$ and $\frac{1}{2}(X - iY)$, respectively. Thus a tensor product of the two acting on qubits $n$ and $n+1$ yields $\frac{1}{4}(X_n X_{n+1} + i Y_n X_{n+1} - i X_n Y_{n+1} + Y_n Y_{n+1})$. This term  is then prefixed with a coefficient $\langle n + 1 | H | n \rangle$. The sum in  Eq.~\autoref{eq:basic_ham} also includes a term with these projectors in the opposite order, yielding $\frac{1}{4}(X_n X_{n+1} - i Y_n X_{n+1} + i X_n Y_{n+1} + Y_n Y_{n+1})$. Due to the symmetry of the deuteron Hamiltonian, $\langle n | H | n + 1 \rangle = \langle n + 1 | H | n \rangle$, and the cross terms cancel leaving $\frac{1}{2} \langle n + 1 | H | n \rangle \left( X_n X_{n+1} + Y_n Y_{n+1} \right) $.

Combining the number- and ladder-type operators, the Hamiltonian in Eq.~\eqref{eq:basic_ham} is re-expressed as
\begin{equation}
H_N=\frac12\sum_{n=0}^{N-1}\braket{n|H|n}(\unit-Z_n)+\frac12\sum_{n=0}^{N-2}\braket{n+1|H|n}(X_{n}X_{n+1}+Y_{n}Y_{n+1}),
\end{equation}
where $\unit$ represents the identity on all qubits.
For $N=2,3$ the Hamiltonian is that given by \cite{Dumitrescu2018} and later \cite{Shehab2019}.  For $N=4$, the Hamiltonian is
 \begin{multline}
 \label{eq:one_hot_4}
 H_4 = 28.657 \unit 
 +0.218 Z_0 
 -6.125 Z_1
 -9.625 Z_2 
 -13.125 Z_3
 -2.143 X_0 X_1\\  
 -3.913 X_1 X_2
 -5.671  X_2 X_3 
 -2.143 Y_0 Y_1 
 -3.913  Y_1 Y_2  
 -5.671  Y_2 Y_3 .
 \end{multline}
 
 \section{Mapping using a Gray code basis ordering}
\label{sec:gray-code}

The one-hot encoding, though simple in implementation, fails to make full use of the qubit states available.  Under the one-hot encoding, the entire $2^N$-dimensional Fock space is mapped onto the $2^N$-dimensional Hilbert space of an $N$-qubit system. However, for particle-conserving operators, like the Hamiltonian, the full set of basis states is not required. For the deuteron example considered here, only $N$ of those states are relevant, and so it can be represented by a system of fewer than $N$ qubits.

The motivation of the Gray code encoding is to perform such a mapping, from an $N$-qubit system using only $N$ of its basis states, down to a $\lceil \log_2 N \rceil$-qubit system using \emph{all} its basis states. In principle, a simple mapping between the states and their binary equivalent will suffice; this case and its disadvantages for this particular problem are discussed in Appendix \ref{appendix:why-not-regular-order}. In this section, it is shown how utilizing a Gray code ordering of the basis states not only uses fewer qubits than a one-hot encoding, but also simplifies the measurement process in the VQE.

\subsection{The Gray code}

A Gray code \cite{gray1953patent} is an ordering of binary values where any two adjacent entries differ by only a single bit. For example,
 \begin{equation} 
    000 \rightarrow \textbf{1}00 \rightarrow 1\textbf{1}0 \rightarrow \textbf{0}10 \rightarrow 01\textbf{1} \rightarrow \textbf{1} 11 \rightarrow 1\textbf{0}1 \rightarrow \textbf{0} 01 \rightarrow 00\textbf{0}.
 \end{equation}
 Gray codes rose to fame in the mid-20th century when they were used for signal conversion in early vacuum-tube televisions \cite{Goodall1951}. They have since found numerous other applications in mathematics, computing, and engineering, such as error correction, Boolean circuit optimization \cite{Karnaugh}, and even quantum circuit synthesis \cite{Vartiainen2004}. 

More formally a Gray code with $\eta$ bits, denoted $\mathbf{G}_\eta$, is given by
\begin{equation}
    \mathbf{G}_\eta = \left( g_0, \enskip g_1, \enskip \ldots, \enskip g_{2^\eta-1} \right),
    \label{eqn:gc_definition}
\end{equation}
where each $g_\alpha$ can be expanded as a sequence of $\eta$ bits $g_\alpha = g_{\alpha,0}, g_{\alpha,1}, \ldots g_{\alpha,\eta-1}$. For example, a Gray code with three bits is \begin{equation}
    \G_3=(000,100,110,010,011,111,101,001).
    \label{eqn:gc_example}
\end{equation}

In addition to the representation as bits, a Gray code $\mathbf{G}_\eta$ can be expressed as a sequence indicating the bit that changes between each step, i.e., 
\begin{equation} 
 \mathbf{S}_{\mathbf{G}_\eta} = \left(s_0, \enskip s_1,  \enskip \ldots,\enskip  s_{2^\eta - 1}\right), \enskip s_i \in  \{0, \enskip  \ldots, \eta - 1\} \enskip  \forall \enskip  i \in 0, \ldots,  2^\eta - 1.
\end{equation}
Explicitly, $s_\alpha = k$ indicates that the $k$th bit is flipped when transitioning from $g_\alpha$ to $g_{\alpha+1}$, with addition in the subscript taken modulo $2^\eta$ as Gray codes are cyclic. For example,  the Gray code in \autoref{eqn:gc_example} can be expressed as
\begin{equation}
    \mathbf{S}_{\mathbf{G}_3} = \left(0, 1, 0, 2, 0, 1, 0, 2 \right).
    \label{eqn:sequence_representation}
\end{equation}

This work makes use of a Gray code construction known as a binary reflective code \footnote{Numerous other constructions for Gray codes exist (for example, \emph{balanced} Gray codes that balance the frequency at which each bit gets flipped \cite{Robinson1981, Bhat1996}). These could be an interesting point of investigation in the context of this work, but are not considered further here.}. For a Gray code $\mathbf{G}_\eta$, let $\overline{\mathbf{G}_\eta}$ represent a Gray code where the $g_\alpha$ appear in the same order but with their bits reversed, i.e., $\overline{g_\alpha} =  g_{\alpha,\eta-1}, g_{\alpha,\eta-2}, \ldots g_{\alpha,0}$. Binary reflective Gray codes are constructed recursively,
\begin{equation}
    \mathbf{G}_\eta = \left( \mathbf{G}_{\eta-1} \cdot 0, \enskip \overline{\mathbf{G}_{\eta-1}} \cdot 1 \right),
\end{equation}
where the center dot indicates concatenation.

\subsection{Gray code basis ordering} 

The essence of the Gray code mapping for Hamiltonians is that a basis state $\ket{n}$ in an $N$-dimensional space is mapped to the multi-qubit state $\ket{g_n}$, $g_n \in \mathbf{G}_\eta$, where $\eta = {\lceil \log_2 N \rceil}$ is the minimum number of qubits required to represent a system with $N$ states. In what follows, $N$ is used to refer to the number of basis states (equivalent to the number of qubits used in the one-hot encoding), while $\eta$ refers to the number of qubits in the Gray code encoding.

For example, if $N=8$ then an example mapping with $\eta = 3$ qubits is
\begin{equation}
\label{eqn:gc_map_3qubit}
 \begin{aligned}
    \ket{0}& \rightarrow \ket{000}, \qquad
    \ket{1} \rightarrow \ket{100}, \qquad
    \ket{2} \rightarrow \ket{110}, \qquad
    \ket{3} \rightarrow \ket{010}, \qquad\\
    \ket{4}& \rightarrow \ket{011}, \qquad
    \ket{5} \rightarrow \ket{111}, \qquad
    \ket{6} \rightarrow \ket{101}, \qquad
    \ket{7} \rightarrow \ket{001}. \qquad
    \end{aligned}
\end{equation}

A new qubit Hamiltonian must now be constructed such that it performs the same action on the qubit states as the  Hamiltonian does on the occupation basis states.
The terms in the deuteron Hamiltonian~\autoref{eq:basic_ham} can be organized into number-operator terms $\ket{n}\bra{n}$ or ladder-operator terms $\ket{n\pm1}\bra{n}$.
 For the Gray code $\mathbf{G}_\eta$ given in \autoref{eqn:gc_definition}, the $\eta$-qubit number operators are defined as
\begin{equation}
 \mathcal{B}_\alpha = \otimes \prod_{\beta = 0}^{\eta - 1} P^{(g_{\alpha, \beta})}, \quad g_\alpha \in \mathbf{G}_\eta \enskip \forall \enskip \alpha = 0, \ldots 2^\eta - 1,
 \label{eqn:gc_number_operators}
\end{equation}
such that $\mathcal{B}_\alpha \ket{g_\alpha} = \ket{g_\alpha}$.

Similarly the ladder operators can be mapped to products of $P^{(i)}$ and $X$ operators.  The form of the ladder operator can be written compactly using the sequence representation of the Gray code, $\mathbf{S}_{\mathbf{G}_\eta}$, given in \autoref{eqn:sequence_representation}. Define the ladder operators
\begin{equation}
    \mathcal{C}_\alpha =  \left( \otimes \prod_{\beta=0}^{s_\alpha - 1} P^{(g_{\alpha,\beta})}\right)  \otimes X \otimes  \left( \otimes \prod_{\beta=s_\alpha + 1}^{\eta -1} P^{(g_{\alpha,\beta})}\right), \quad s_\alpha \in \mathbf{S}_{\mathbf{G}_\eta},
    \label{eqn:gc_ladder_operators}
\end{equation}
such that $\mathcal{C}_\alpha \ket{g_\alpha} = \ket{g_{\alpha+1}}$. Intuitively, this applies an $X$ on the qubit $s_\alpha$ that indicates the flipped bit in the Gray code between $g_\alpha$ and $g_{\alpha+1}$; the remaining qubits are kept in their present state using the associated projectors. 

By replacing the operators in \autoref{eq:basic_ham} by those of \autoref{eqn:gc_number_operators} and \autoref{eqn:gc_ladder_operators}, the full qubit Hamiltonian for the deuteron under the Gray code encoding is
\begin{equation}
H_N= \sum_{\alpha=0}^{2^\eta -1}\braket{g_\alpha|H|g_\alpha} \mathcal{B}_\alpha  +  \sum_{\alpha=0}^{2^\eta-2}\braket{g_{\alpha + 1}|H|g_\alpha} \mathcal{C}_\alpha,
\label{eqn:gc_hamiltonian}
\end{equation}
where $\eta = \left\lceil \log_2 N \right\rceil$ and $ g_\alpha \in \mathbf{G}_\eta$. Under the mapping of $\ket{n} \rightarrow \ket{g_n}$, the matrix elements $\braket{g_{n + 1}|H|g_n} = \braket{n + 1|H|n} = \braket{n|H|n+1}$ for $n = 0, \ldots, N - 1$, which are given in \autoref{eq:matel}.
 
\subsection{Examples\label{subsec:gc_examples}}

Unlike the one-hot encoding which mapped the Hamiltonian with $N=4$ to four qubits in \autoref{eq:basic_ham}, the Gray code encoding can map this same Hamiltonian onto only two qubits, for example
 \begin{equation}
 \label{eqn:gc_map_2qubit}
     \ket{0} \rightarrow \ket{00}, \qquad
     \ket{1} \rightarrow \ket{10}, \qquad
     \ket{2} \rightarrow \ket{11}, \qquad
     \ket{3} \rightarrow \ket{01}. 
 \end{equation}
 The corresponding Hamiltonian is then given by \autoref{eqn:gc_hamiltonian} for a two-qubit system.  The specific number and ladder operators for this mapping are summarized in Table~\ref{tab:gc_2qubit} in  Appendix \ref{appendix:encoding-tables}. The resulting Hamiltonian is given by 
 \begin{multline}
 H_4 = \textstyle{\frac{1}{4}}\left[\braket{0|H|0} + \braket{1|H|1} + \braket{2|H|2} + \braket{3|H|3}\right] \unit 
 + \textstyle{\frac{1}{2}}\left[\braket{0|H|1} + \braket{2|H|3} \right]  X_0 \\
 + \textstyle{\frac{1}{2}} \braket{1|H|2} X_1 
 + \textstyle{\frac{1}{4}}\left[\braket{0|H|0} - \braket{1|H|1} - \braket{2|H|2} + \braket{3|H|3}\right]  Z_0 \\
 + \textstyle{\frac{1}{4}}\left[\braket{0|H|0} + \braket{1|H|1} - \braket{2|H|2} - \braket{3|H|3} \right]  Z_1
 + \textstyle{\frac{1}{2}}\left[\braket{0|H|1} - \braket{2|H|3} \right]X_0 Z_1 \\
 - \textstyle{\frac{1}{2}} \langle 1|H|2\rangle  Z_0 X_1 
 + \textstyle{\frac{1}{4}}\left[\braket{0|H|0} - \braket{1|H|1} + \braket{2|H|2} - \braket{3|H|3}\right]  Z_0 Z_1 \; ,
 \end{multline}
 where the matrix elements are as expressed in \autoref{eq:matel}. Evaluating these elements with the selected parameter values gives
 \begin{multline}
 \label{eq:gray_code_4}
 H_4 = 14.328\unit
 -7.814 X_0-3.913 X_1
 -1.422  Z_0-8.422  Z_1
 +3.527 X_0 Z_1
 +3.913 Z_0 X_1
 -4.922 Z_0 Z_1. 
 \end{multline}
As the Gray code encoding for two qubits is particularly simple, it is instructive to consider an example with more qubits.  As shown in \autoref{eqn:gc_map_3qubit}, the eight states that make up the $N=8$ basis can be mapped onto only three qubits.  Evaluating \autoref{eqn:gc_hamiltonian} for this case yields  

\begin{multline}
H_8=29.039\unit
- 0.711 Z_0 - 0.711 Z_1 -14.711 Z_2 \\
- 0.711 Z_0 Z_1 - 0.711  Z_0 Z_2 - 7.711 Z_1 Z_2 - 4.211 Z_0 Z_1 Z_2\\
 - 14.835 X_0 + 0.012 X_0 Z_1  + 7.022 X_0 Z_2  + 3.515 X_0 Z_1 Z_2   \\
- 7.421 X_1 + 7.421 Z_0 X_1  + 3.508 X_1 Z_2 - 3.508 Z_0X_1Z_2\\
-3.712 X_2  -3.712  Z_0 X_2 + 3.712  Z_1 X_2 + 3.712 Z_0 Z_1 X_2.\\
\end{multline}
The specific form of the number and ladder operators in the Gray code ordering used to obtain the Hamiltonian above from the general expression \autoref{eqn:gc_hamiltonian} for $N=8$ are summarized in \autoref{tab:gc_3qubit} of Appendix \ref{appendix:encoding-tables}.

\subsection{Pauli structure of the Hamiltonians}
\label{subsec:paulistructure}

An important consideration for both the VQE and simulating time evolution is the way in which the constituent Paulis of a Hamiltonian can be partitioned into commuting sets. In the VQE, this partitioning reduces the number of measurements that need to be taken, since expectation values of commuting Paulis can be measured simultaneously. For simulating time evolution, the structure of the commuting sets and the order in which Paulis are written affects the accuracy of the simulation. These aspects will be discussed further in \autoref{subsec:vqe} and \autoref{sec:hamsim}, respectively.

A key feature of the one-hot encoding is that Pauli terms in the Hamiltonian partition into the three commuting sets shown in \autoref{tab:oh_pauli_sets}. Note that the weight of the Pauli strings is at most 2. In addition, within each set, measurements on all qubits take place in the same basis so there is no need to perform a rotation to a common eigenbasis when measuring the expectation values of these Pauli operators; we need only apply a Hadamard $H$ to all qubits for the set with $X$, or $H S^\dag$ for the set with $Y$, where 
\begin{equation}
    H = \frac{1}{\sqrt{2}} \begin{pmatrix}
     1 & 1 \\
     1 & -1 
    \end{pmatrix}, 
    \quad 
    S = \begin{pmatrix}
     1 & 0 \\
     0 & i 
    \end{pmatrix}.
\end{equation}

\begin{table}[ht]
    \caption{Structure of the sets of commuting operators for the deuteron Hamiltonian using the one-hot encoding.}
    \centering
    \begin{tabular}{c|c}
         0 & $Z_0, \enskip Z_1,\ldots, \enskip Z_{N-1}$ \\
         1 & $Y_0 Y_1, \enskip Y_1 Y_2, \ldots, Y_{N-2} Y_{N-1}$\\
         2 & $X_0 X_1, \enskip X_1 X_2,  \ldots, X_{N-2} X_{N-1}$\\
         \end{tabular}
    \label{tab:oh_pauli_sets}
\end{table}

From the examples in \autoref{subsec:gc_examples} one observes that a consequence of the Gray code encoding is that each $\eta$-qubit Pauli term contains at most a single $X$. This naturally partitions the operators into $\eta+1$ commuting sets: one containing all combinations of $\unit$ and $Z$, and the rest containing an $X$ on a given qubit, and then all combinations of $\unit$ and $Z$ on the rest.  For the $\eta = 3$ case, there are four sets:
\begin{equation}
 \begin{aligned}
S_{Z} &= \{ Z_0, \enskip Z_1, \enskip Z_2, \enskip  Z_0 Z_1, \enskip  Z_1 Z_2, \enskip Z_0 Z_2, \enskip  Z_0 Z_1 Z_2 \}, \\ 
S_{X0} &= \{X_0, \enskip X_0 Z_1, \enskip  X_0 Z_2, \enskip  X_0 Z_1 Z_2 \}, \\ 
S_{X1} &= \{X_1, \enskip Z_0 X_1, \enskip X_1 Z_2, \enskip  Z_0 X_1 Z_2 \}, \\ 
S_{X2} &= \{X_2, \enskip Z_0 X_2, \enskip Z_1 X_2, \enskip  Z_0 Z_1 X_2 \}. 
 \end{aligned}
\end{equation}
Thus, for each commuting set, only one qubit ever needs to be rotated back to the computational basis. This reduces the number of such rotations over measurement of all sets to $\eta$, down from $3N$ in the one-hot encoding.

It is straightforward to generalize the structure of the Hamiltonians to arbitrary $N$. When $N$ is not a power of 2, a truncated Gray code can be used, but will still require $\eta = \lceil \log_2 N \rceil$ qubits. \autoref{tab:hamiltonian_comparison} compares key properties of the Hamiltonians of the two encodings for an $N$-state problem.

\begin{table}[ht]
    \caption{Comparison of Hamiltonian encodings of an $N$ state system.}
    \centering
    \setstretch{1.2}
    \begin{tabular}{|l|c|c|}
     \hline
     Encoding & One-hot & Gray code  \\\hline \hline
     Qubits & $N$ & $\lceil \log_2 N \rceil = \eta $  \\ \hline
     Number of Pauli terms & $3 N - 2$ & $2^\eta + \eta 2^{\eta-1} - 1$ \\ \hline
     Commuting sets of Paulis & 3 & $\eta+1$ \\ \hline
     Max Pauli weight & 2 & $\eta$ \\ \hline
    \end{tabular}
    \label{tab:hamiltonian_comparison}
\end{table}

The structure of the Hamiltonians reveals an interesting trade-off.  For the same number of states, the Gray code Hamiltonian has more terms, and higher weight of the Pauli strings per term. Furthermore, the number of commuting sets is no longer constant, which may have consequences while running the VQE as more measurements must be made. However, the number of qubits is exponentially smaller, so it is necessary to explore whether the trade-off of using fewer qubits is beneficial, especially in a noisy hardware setting.

\section{Finding the ground state energy with the variational quantum eigensolver}
\label{sec:implementation}

\subsection{Variational quantum eigensolver}
\label{subsec:vqe}
The VQE \cite{Peruzzo2014} is based on the variational principle: given a Hamiltonian $H$ with ground state $\ket{\psi_g}$ and energy $E_g$, the expectation value for any other state will always be greater,
\begin{equation}
    \bra{\psi} H \ket{\psi} \geq \bra{\psi_g} H \ket{\psi_g} = E_g.
\end{equation}
The VQE parameterizes the state $\ket{\psi}$ as $\ket{\psi(\theta)}$, and uses classical optimization to search for a set of suitable parameters $\theta$ such that
\begin{equation}
    \bra{\psi(\theta)} H \ket{\psi(\theta)} = E_g.
\end{equation}
In practice, the state $\ket{\psi(\theta)}$ is expressed as the action of a variational ansatz circuit acting on an initial state, typically $\ket{0}$: $\ket{\psi(\theta)} = U(\theta) \ket{0}$. The goal is then to find a suitable ansatz $U(\theta)$ and its parameters such that the resultant expectation value is as close as possible to the true ground state of the system. 
A Hamiltonian on $N$ qubits, being a Hermitian matrix, can be expressed as a linear combination of the $N$-qubit Pauli operators,
\begin{equation}
    H = \sum_{i=0}^{4^N - 1} q_i Q_i, \quad Q_i \in \mathcal{P}_N,
    \label{eqn:ham_sum_paulis}
\end{equation} 
for expansion coefficients $q_i \in \mathds{R}$. The expectation value of the Hamiltonian can be computed as a linear combination of the expectation values of the individual terms,
\begin{equation}
    \langle H \rangle = \sum_{i=0}^{4^N - 1} q_i \langle Q_i \rangle, \quad Q_i \in \mathcal{P}_N.
\end{equation}
The role of the quantum computer is to apply $U(\theta)$ and take measurements to obtain these expectation values, which are then processed on a classical computer during an optimization routine. Based on the results, a new value of $\theta$ is chosen and the process is repeated again. To reduce the number of measurements, one typically takes advantage of the fact that expectation values of sets of Paulis that commute can be measured simultaneously. Algorithms for creating and analyzing such sets are under active development \cite{Gokhale2019, jena2019pauli, gokhale2019on3, verteletskyi2019measurement}; the upper bound for an $\eta$-qubit system is $2^\eta + 1$, which corresponds to measuring in a complete set of mutually unbiased bases.

\subsection{Choosing a variational ansatz}

While it is always possible to apply a generic hardware-efficient variational ansatz \cite{Kandala2017}, it is beneficial to choose an ansatz  informed by the problem at hand.
The nuclear physics problem considered here is defined by a real Hamiltonian matrix which is diagonalizable by an \emph{orthogonal} transformation.
As a consequence the most general unitary-state preparation is not necessary, and a more targeted ansatz can be applied for both encodings.

For the one-hot encoding, a variational form with coefficients that are real functions of generalized spherical coordinates will be able to access the entire Hilbert space spanned by the desired basis vectors.
For $N$ states this requires $N$ qubits and $N-1$ parameters. 
As an example, for the 4-state case
\begin{equation} 
 \ket{\psi} = \cos \theta_1  \ket{0001} +  \sin \theta_1  \cos \theta_2 \ket{0010} +   
 \sin \theta_1 \sin \theta_2  \cos \theta_3 \ket{0100} +  \sin \theta_1 \sin \theta_2 \sin \theta_3 \ket{1000}.
 \label{eq:oh_wavefunction}
\end{equation}

\begin{figure}[ht]
    \centering
    \includegraphics[scale=2]{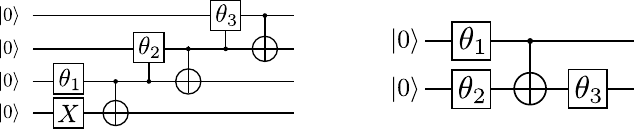}
        \caption{Variational ansatz for $N=4$ used with the one-hot encoding (left) and the Gray code encoding (right). The gates indicated by $\theta_i$ are Pauli $Y$ rotations. }
    \label{fig:ansatze-4state}
\end{figure}

Such states can be constructed recursively using a cascade of controlled rotations and CNOTs \cite{Shehab2019}. The circuit for the 4-qubit case is shown in \autoref{fig:ansatze-4state}(left). Note that the ordering of the qubits here is reversed from that of~\cite{Shehab2019}, so that the basis state with the largest contribution to the ground state depends on all parameters. This reordering was found to provide improved stability during the optimization procedure when using the state vector simulator.

\begin{figure}[ht]
    \centering
    \includegraphics[scale=1]{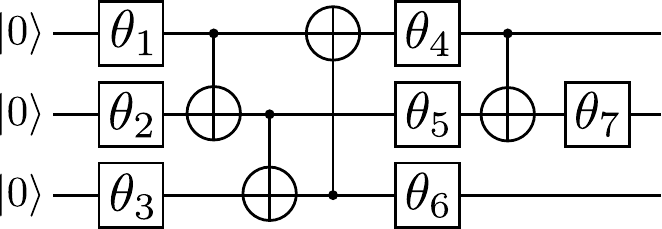}
        \caption{Variational ansatz for $N=8$ using the Gray code encoding.}
    \label{fig:ansatz-8state}
\end{figure}

The Gray code encoding, since it incorporates all available states, enables the use of  a streamlined hardware-efficient variational ansatz consisting of layers of parametrized $Y$ rotations separated by layers of entangling gates \cite{Kandala2017}. The right panel of \autoref{fig:ansatze-4state} presents such an ansatz for the $N=4$ case using only two qubits; a further example with $N=8$ is shown in \autoref{fig:ansatz-8state}. The wave function for the $N=4$ case can be evaluated as
\begin{equation}
\begin{split}
    \ket{\psi} = & \cos\theta_1 \cos (\theta_2 + \theta_3) \ket{00} + 
    \sin\theta_1 \sin (\theta_2 - \theta_3) \ket{10} + \\ 
    & \sin\theta_1 \cos (\theta_2 - \theta_3) \ket{11} + 
    \cos\theta_1 \sin (\theta_2 + \theta_3) \ket{01}.
\end{split}
\end{equation}
In contrast to the wave function of \autoref{eq:oh_wavefunction}, the coefficient for each basis state in the wavefunction above depends on every variational parameter.

The structure of the variational ansatz has a significant effect on the success of the VQE, especially in a noisy hardware environment. The number of gates, circuit depth, and in particular number of two-qubit gates are all important points of comparison. In what follows, the resources required to run the VQE ansatz of an $N$-state problem are computed, with results summarized in \autoref{tab:circuit_comparison}.

To estimate the resource requirements for the one-hot encoding ansatz of \autoref{fig:ansatze-4state}, the controlled rotations are first decomposed into two single-qubit rotations and two CNOTs. Thus, an $N$-qubit version of this circuit uses $3N - 5$ two-qubit gates and $2N - 2$ single-qubit gates, and runs in depth $4N - 6$. 
Additional single-qubit rotations must also be performed to rotate back to the computational basis when measuring the commuting Pauli sets with $X$ and $Y$ (as was described in \autoref{subsec:paulistructure}). These require $N$ and $2N$ additional rotations respectively. Considering the execution of circuits for all three commuting sets of Paulis, the total number of gates is $18N - 21$.
 
 For the Gray code encoding ansatz, since there are no controlled rotations to decompose, the number of single-qubit gates depends on the number of qubits $\eta = \lceil \log_2 N \rceil$. There are $2^\eta - 1$ single-qubit gates, and $2^\eta - 1 - \eta$ two-qubit gates. This can be seen from the structure of the circuit as alternating layers of single-qubit gates and CNOTs; each single-qubit gate is paired with a CNOT to its left, save for the first layer. Calculation of the depth is slightly more involved, but is shown in \autoref{tab:circuit_comparison}. Finally, the structure of the commuting Pauli sets leads to a simple measurement procedure. To measure the set $S_{Xi}$, one simply performs a Hadamard on qubit $i$ before measurement to rotate it back to the computational basis. Across all sets this yields only $\eta$ extra single-qubit gates, as opposed to $3N$.
 
\begin{table}[ht]
    \caption{Comparison of variational ansatz circuits for a Hamiltonian with $N$ states for the two encodings. Numbers shown are the gate counts for a single evaluation of the expectation value using VQE, i.e., measuring all commuting sets of Paulis. The depth is given for the circuit without any additional basis rotations.}
    \centering
    \setstretch{1.2}
    \begin{tabular}{|l|c|c|}
     \hline
     Encoding & One-hot & Gray code  \\\hline \hline
       Qubits     & $N$ & $ \lceil \log_2 N \rceil = \eta$ \\ \hline
       Single-qubit gates & $3(2 N - 2)$ & $(\eta + 1) \left(2^\eta - 1 \right)$ \\ \hline
       Additional basis rotations & $3N$  & $\eta$  \\ \hline
       Two-qubit gates & $3 (3 N - 5)$ & $(\eta + 1) \left(2^\eta - \eta - 1 \right)$ \\ \hline \hline
       Total gates for VQE & $18 N - 21$ & $2(\eta+1)(2^\eta-1) - \eta^2$ \\ \hline \hline
       Individual circuit depth   &  $4N - 6$ & $\left\lceil \frac{2^\eta-1}{\eta} \right\rceil (\eta + 1) - 2\eta + (2^\eta-1)\hbox{mod}\eta$ \\ \hline
    \end{tabular}
    \label{tab:circuit_comparison}
\end{table}

\autoref{tab:circuit_comparison} shows that the depth of the Gray code ansatz is consistently better than the one-hot ansatz. One interesting point, however, is that the number of two-qubit (and total) gates in the Gray code ansatz always surpasses that of the ansatz in \cite{Shehab2019} starting at $N = 256$ (for lower $N$, the Gray code gate counts are lower for $N$ that are powers of 2). However one can surely argue that if one has access to that many qubits, we are no longer in the NISQ era and there are better methods available than the VQE.


\subsection{VQE simulation results}
\label{subsec:vqe-results}

The quantum computing component of the implementation was simulated using Qiskit \cite{Qiskit} and OpenFermion \cite{OpenFermion}. The simultaneous perturbation stochastic approximation (SPSA) algorithm \cite{Spall1992, Spall1998, Spall1999}  was chosen as the classical optimization routine, using the implementation provided in the \texttt{noisyopt} Python package ~\cite{NoisyOpt}. SPSA was run using step parameters $a=0.628, c=0.1$. Initial values for the variational parameters were chosen uniformly at random from the range $(-\pi/2, \pi/2)$.  The number of iterations used was 2000 for $N=2$, 4000 for $N=3$, 5000 for $N = 4, \ldots, 8$ and 8000 for $N = 16$. Simulations for $N=16$ were performed only with the Gray code encoding, due to the computational intensity of simulating thousands of VQE steps for a 16-qubit system. The implementation, as well as the data files and initial parameters, is available at \cite{OurCode}.

The encoding was first analyzed using Qiskit's state vector simulator to verify correctness. This was followed by testing with the QASM (quantum assembly language) simulator, which simulates the probabilistic behavior of quantum computers and returns counts of the different measurement outcomes rather than an analytical solution. These simulations were run with 10000 trials (``shots") per circuit. For each $N$,  100 independent trials of the full VQE were performed. Since the number of iterations of SPSA is fixed, the quantity of interest is the solution quality and variance as compared to the true value obtained from diagonalization.

\begin{figure}
\centering
    \begin{tabular}{c}
        \includegraphics[width=.5\textwidth]{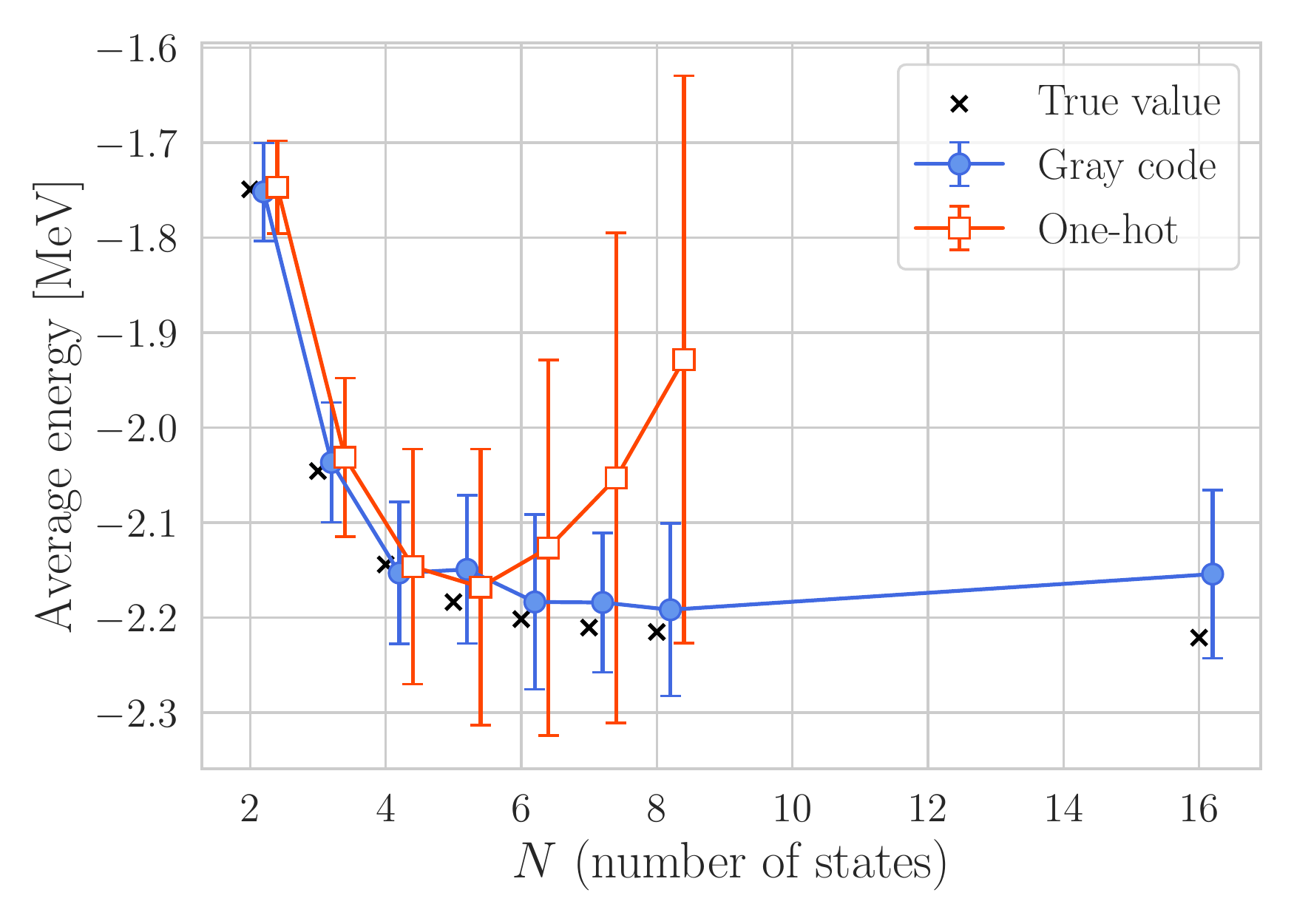}
        \includegraphics[width=.5\textwidth]{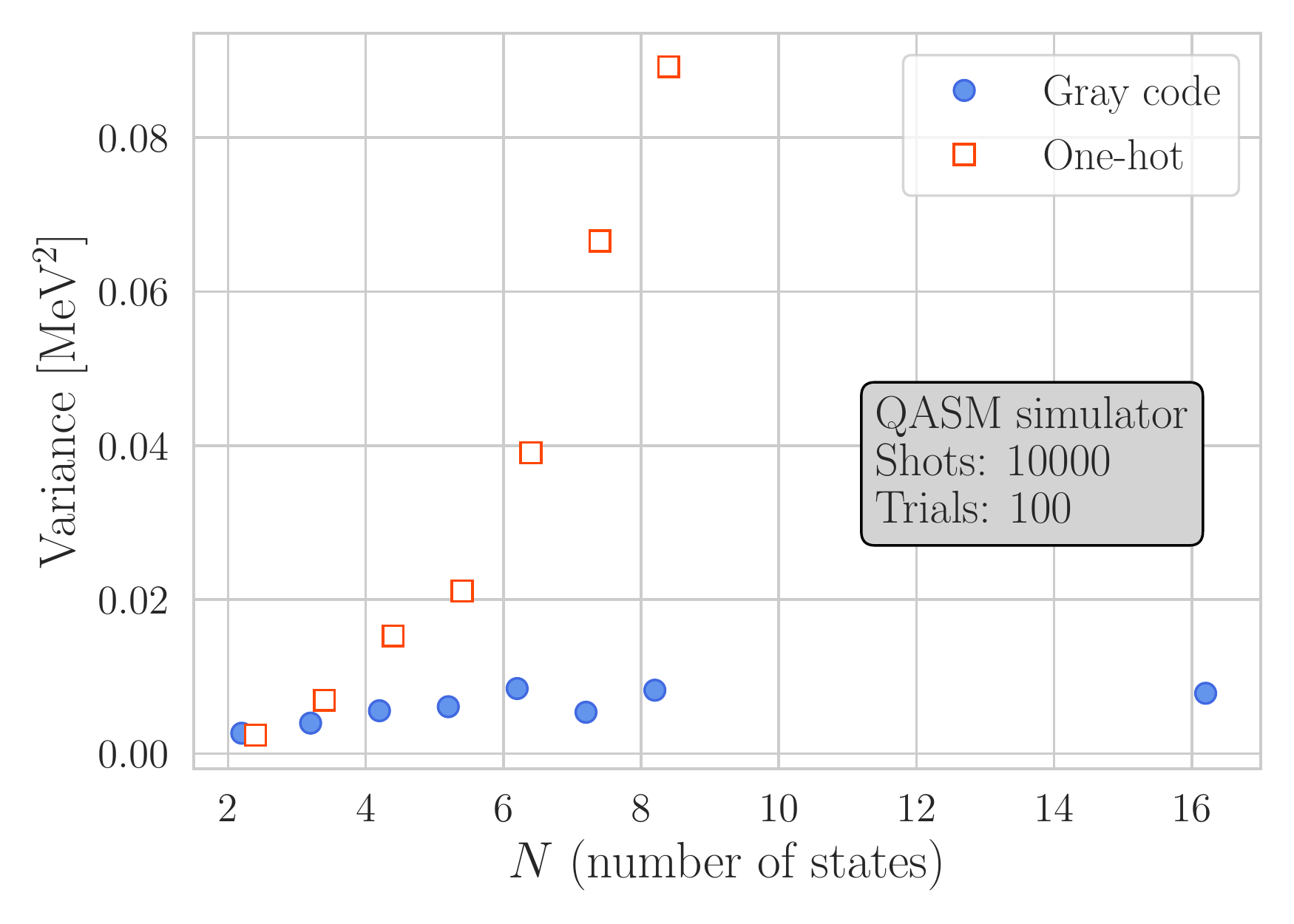} \\
    \end{tabular}
\caption{Comparison of the average energy obtained over 100 independent executions of the VQE using the Qiskit QASM simulator with 10000 shots. Error bars in the left panel show the standard deviation, and values are offset for clarity. There is significantly more variance in the energies obtained for the one-hot encoding. This is interesting as it shows that, despite having more Pauli terms and more commuting sets to measure in the Gray code case, the results are consistently closer to the true value in the presence of shot noise.}
\label{fig:qasm-results}
\end{figure}

\begin{figure}
\centering
    \begin{tabular}{c}
        \includegraphics[width=0.5\linewidth]{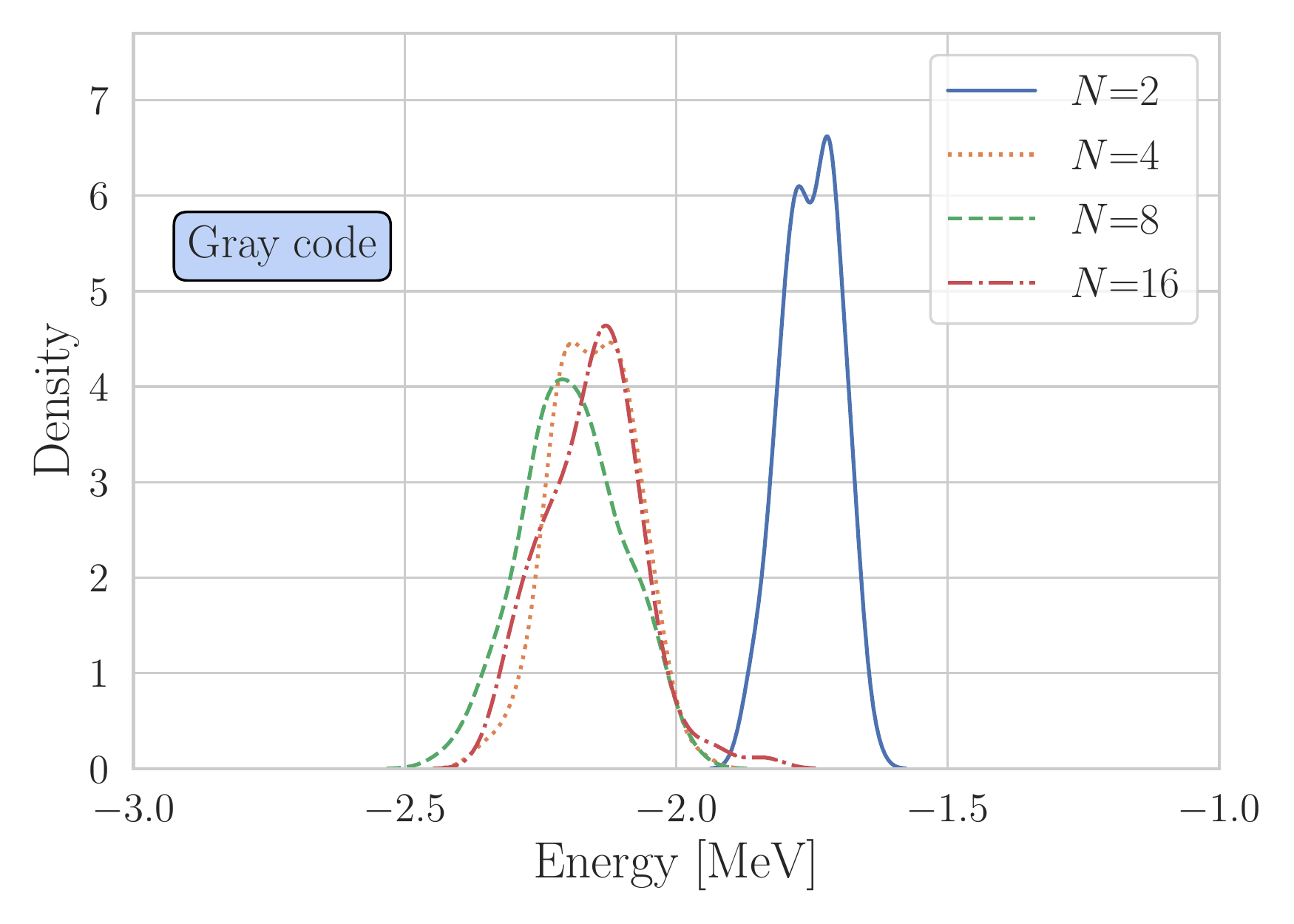} 
        \includegraphics[width=0.5\linewidth]{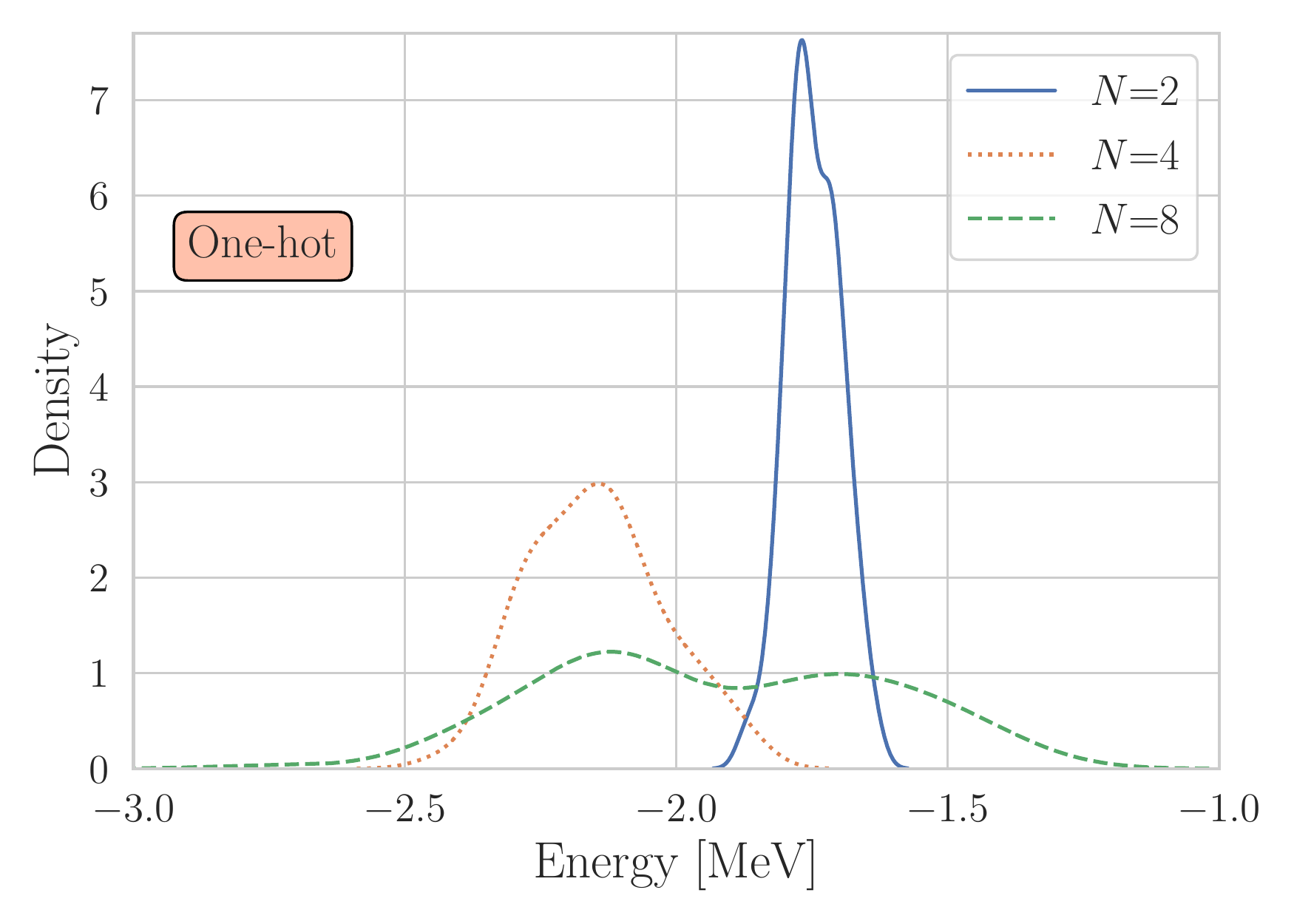} \\
    \end{tabular}
\caption{The data from \autoref{fig:qasm-results} as a density plot. There is significantly more spread of values for the one-hot encoding (see right panel of \autoref{fig:qasm-results} for the explicit variances).}
\label{fig:density-plots}
\end{figure}

Results for the QASM simulations are plotted in \autoref{fig:qasm-results}. The variance of solutions is observed to be significantly higher for the one-hot encoding. 
This difference is even more visible in \autoref{fig:density-plots}, the density plots of the energy distribution for the QASM simulations, and the effect is amplified as $N$ increases. 
One might think that the small variance in the Gray code simulation is attributed to the SPSA minimization error, or a consequence of the structure of the wave function ansatz.
To check this, we performed the same simulation again with the optimized angles obtained with the exact state vector simulation.
We observed exactly the same behavior shown in the right panel in Fig.~\ref{fig:qasm-results}, and thus the SPSA minimization algorithm is not the source of behavior of the variance.

To further investigate, we computed the covariance matrix of Pauli operators for the two types of Hamiltonians.
Principal component analysis revealed that the number of effective degrees of freedom with the Gray code encoding is smaller than that with the one-hot.
Combined with the fact that the sizes of the coefficients in front of each Pauli term are the same order of magnitude in the one-hot and Gray code encodings, see Eqs.~\eqref{eq:one_hot_4} and~\eqref{eq:gray_code_4} for example, the small variance of the Gray code results is understandable as a result of structural differences in the Hamiltonians.
This demonstrates that the benefits of using a different encoding extend beyond the obvious advantages of a reduced number of qubits, or favorable structure of the ansatz; the form of the Hamiltonian itself plays a critical role.

\subsection{Resilience in the presence of simulated hardware noise}
\label{subsec:hardware-noise}

\begin{figure}
    \begin{centering}
        \includegraphics[width=0.5\textwidth]{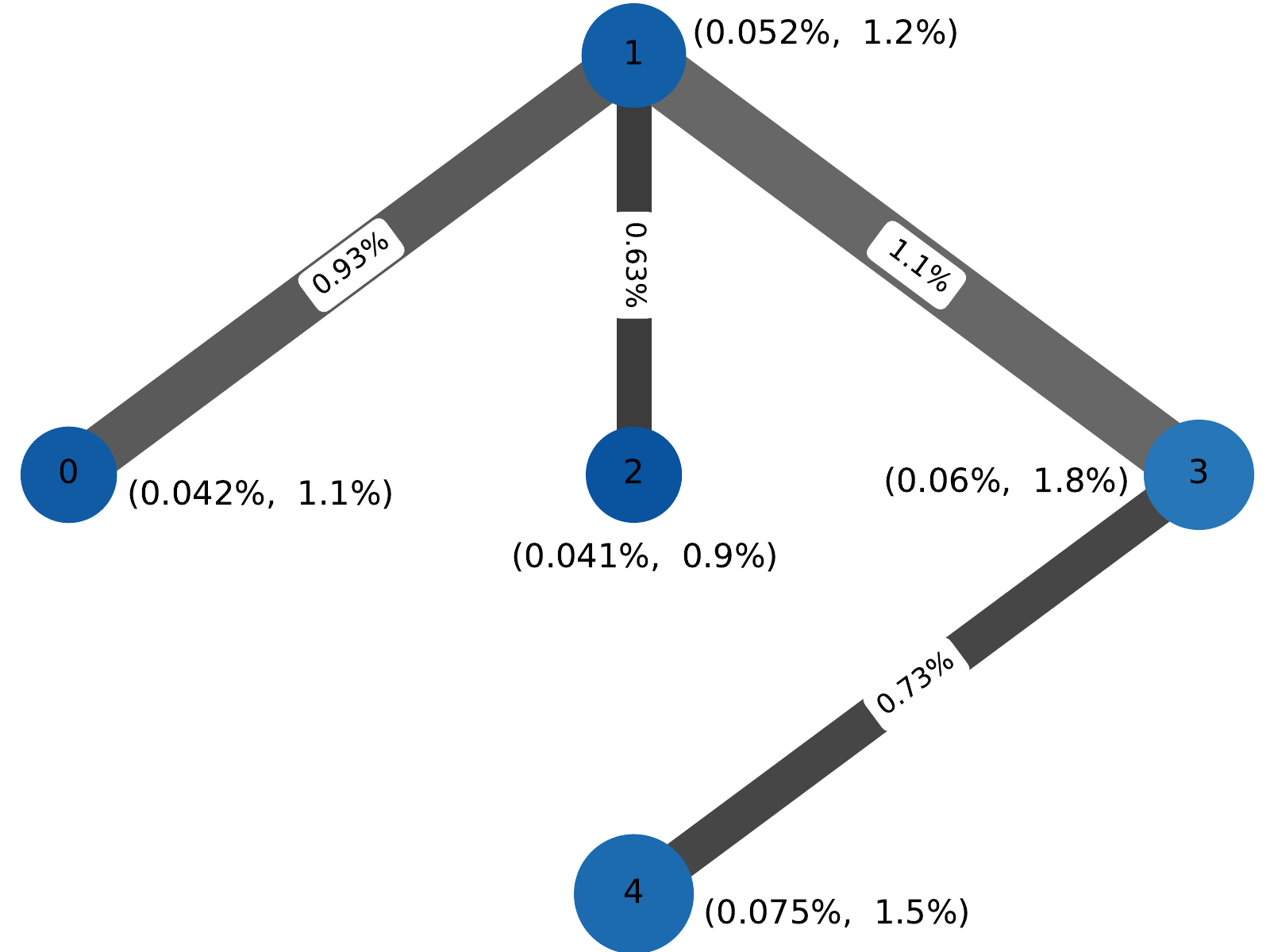}  
    \end{centering}
    \caption{Hardware graph for the IBM Q machine Vigo. Each node of the graph corresponds to a physical qubit. Calibration data were retrieved from the IBM Q Experience portal on July 8, 2020. The pair of values in the node label corresponds to the single-qubit gate error rate (left) and measurement error rate (right) that were used for the simulations. The edge label corresponds to the two-qubit gate error rate. Lighter color, larger node size, and larger edge width correspond to higher error rates.}
    \label{fig:hardware_graphs}
\end{figure}

\begin{figure}
    \begin{tabular}{c}
        \includegraphics[width=.5\textwidth]{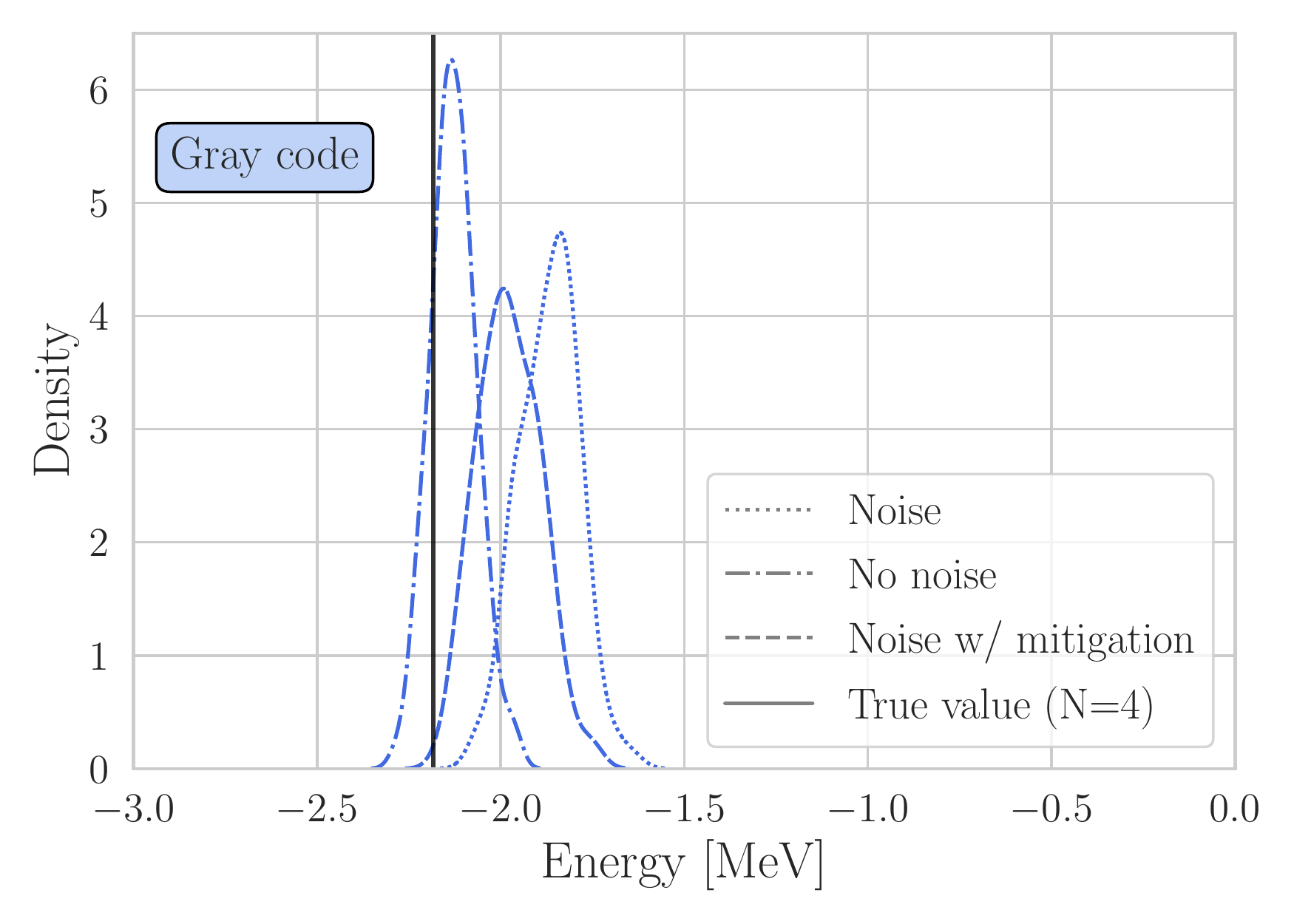}
        \includegraphics[width=.5\textwidth]{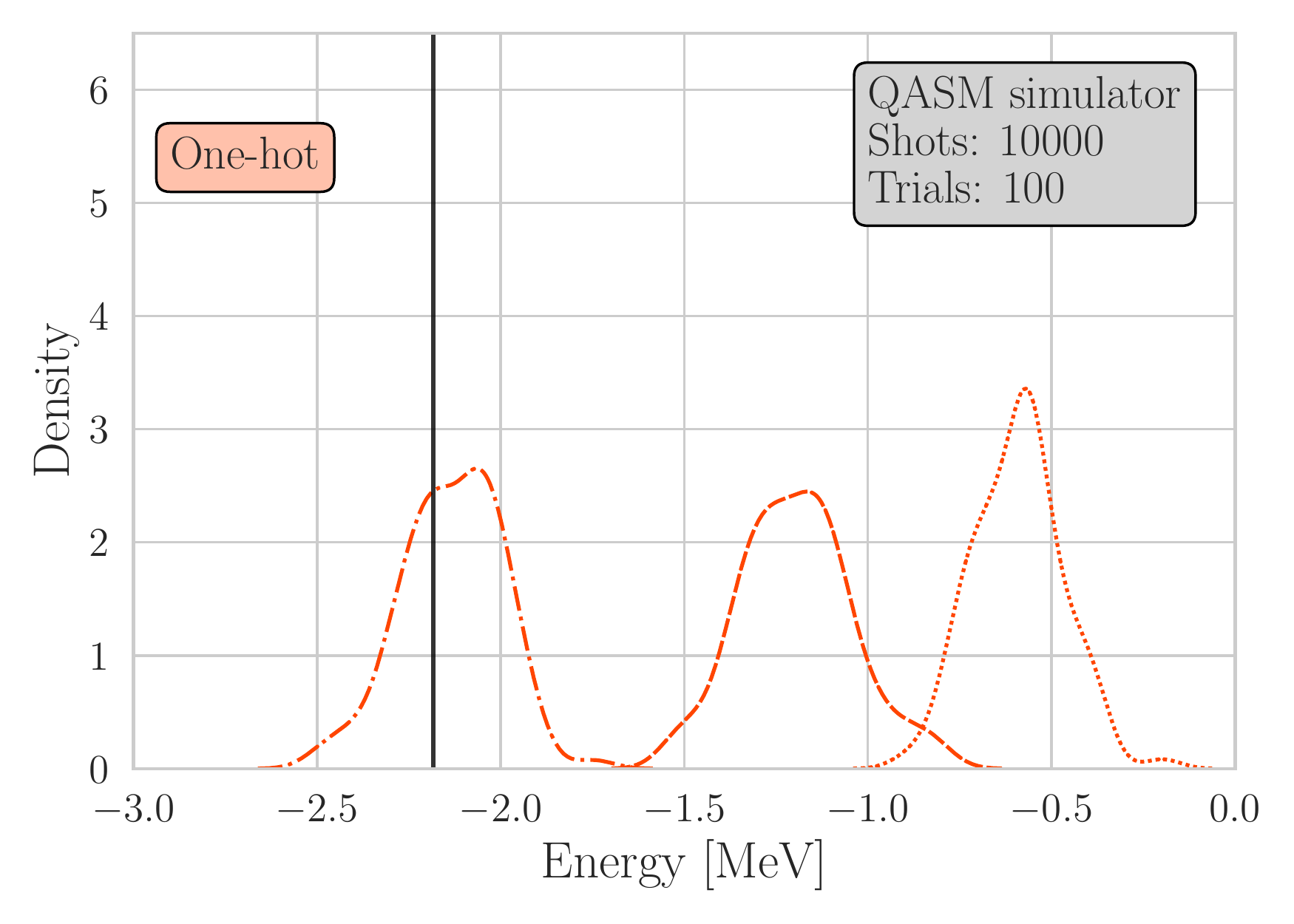} \\
    \end{tabular}
    \caption{Distribution of VQE energies in QASM simulations (10000 shots) with additional simulated hardware noise using the Vigo device noise model, noise with measurement error mitigation, and no simulated hardware noise. Each simulation is carried out using a particular layout of qubits on the hardware graph. The simulation using the Gray code maps the logical qubits \{0,1\} to the ``physical'' qubits \{2,1\} of the layout shown in \autoref{fig:hardware_graphs}. The one-hot simulation maps logical qubits \{0,1,2,3\} to physical qubits \{2,1,3,4\}.}
    \label{fig:noise_comparison}
\end{figure}

Given that the Gray code encoding uses fewer qubits and has circuits of shorter depth with fewer two-qubit gates, it is reasonable to expect its performance may improve over that of the one-hot encoding when there is hardware noise present.
To investigate this, a noise model from the IBM Q device Vigo \footnote{\emph{ibmq\_vigo} v1.0.2, IBM Quantum team. Retrieved from https://quantum-computing.ibm.com (2020)} was applied, and simulations for both encodings are compared in the $N=4$ case.
Results from an additional simulated IBM device are shown in Appendix \ref{appendix:additional-noise-models}.
The hardware graph and error rates for the simulated device are shown in \autoref{fig:hardware_graphs}. 
The noise model approximates the physical device by implementing single-qubit and two-qubit gate errors as well as measurement readout errors.
Data for the particular noise models used are provided in the code \cite{OurCode}.  The results of the simulations are shown in \autoref{fig:noise_comparison}.

Unsurprisingly, the one-hot version suffers far more from the additional noise than the Gray code version.
Even after performing measurement error mitigation, i.e., correcting for expected errors based on calibration circuits (carried out using Qiskit's Ignis library), the obtained results are displaced from the exact answer, which is calculable on classical computers.
The Gray code is much closer, both before and after mitigation, with some overlap of the distribution with the exact value.
These findings suggest that in the near term, it may be beneficial to use an encoding with fewer qubits and shorter circuits despite the trade-offs in the structure of the Hamiltonian, i.e., needing to make far more measurements.

To obtain results on noisy hardware that are comparable to those from the clean simulations, extrapolation to the noiseless limit can be performed using a technique called zero-noise extrapolation \cite{Li2017,Temme2017,Kandala_2019}.
Let $E(\varepsilon)$ be an expectation value depending on a noise parameter $\varepsilon$.
The noiseless result is $E(\varepsilon=0)$ and the simulation result is $E(\varepsilon=\varepsilon_{0})$, where $\varepsilon_{0}$ is the noise parameter of a NISQ device.
The value of $E(\varepsilon=0)$ can be estimated by simulating $E(\varepsilon)$ at $\varepsilon > \varepsilon_{0}$ and extrapolating to $\varepsilon=0$.
Several discussions have been made to calculate $E(\varepsilon)$ at $\varepsilon$ other than $\varepsilon_{0}$~\cite{Li2017,Temme2017,Dumitrescu2018,Shehab2019,Kandala_2019,He2020,UnitaryFolding}.
The method used in this work is to add redundant CNOT gate pairs to the original circuit to simulate $E(\varepsilon > \varepsilon_{0})$ as done in the earlier deuteron simulations~\cite{Dumitrescu2018,Shehab2019}.
Since the noise of a single-qubit gate is much smaller than that of two-qubit gates (CNOT gate), $E(\varepsilon)$ is mainly affected by the number of CNOT operations.
This suggests that $\varepsilon$ can be scaled by the number of CNOTs in the circuit; e.g., $\varepsilon$ would be $2\varepsilon_{0}$ when the number of CNOTs is twice that of the original circuit~\cite{He2020}.
Given that $(\hbox{CNOT})^{2n}= 1$ is satisfied at the $\varepsilon_{0}=0$ limit (for $n=0,1,2,...$), $E(\varepsilon)$ at $\varepsilon=(2n+1)\varepsilon_{0}$ can be calculated by replacing every CNOT gate in the original circuit with $(\hbox{CNOT})^{2n+1}$. (The choice of the qubit pair would also affect the value of $\varepsilon$.)
Expanding $E(\varepsilon)$ around $\varepsilon=0$ and plugging in $\varepsilon=(2n+1)\varepsilon_{0}$, $E(\varepsilon)$ is
\begin{equation}
\label{eq:extrap}
E(\varepsilon)  \sim E(0) +   \varepsilon_{0} \left. \frac{dE(\varepsilon)}{d\varepsilon}\right|_{\varepsilon=0} (2n+1).
\end{equation}
Note that $\varepsilon_{0}\sim m \times 10^{-2}$, using the number of CNOT operators in the original circuit $m$ (see Table~\ref{tab:hamiltonian_comparison} and Fig.~\ref{fig:hardware_graphs} for practical cases), on currently available devices, and higher-order terms in $\varepsilon_{0}$ should be negligible.
Eq.~(\ref{eq:extrap}) enables us to extrapolate to the zero-noise limit in terms of $(2n+1)$ instead of $\varepsilon$.

\begin{figure}
    \begin{tabular}{c}
        \includegraphics[width=0.5\textwidth]{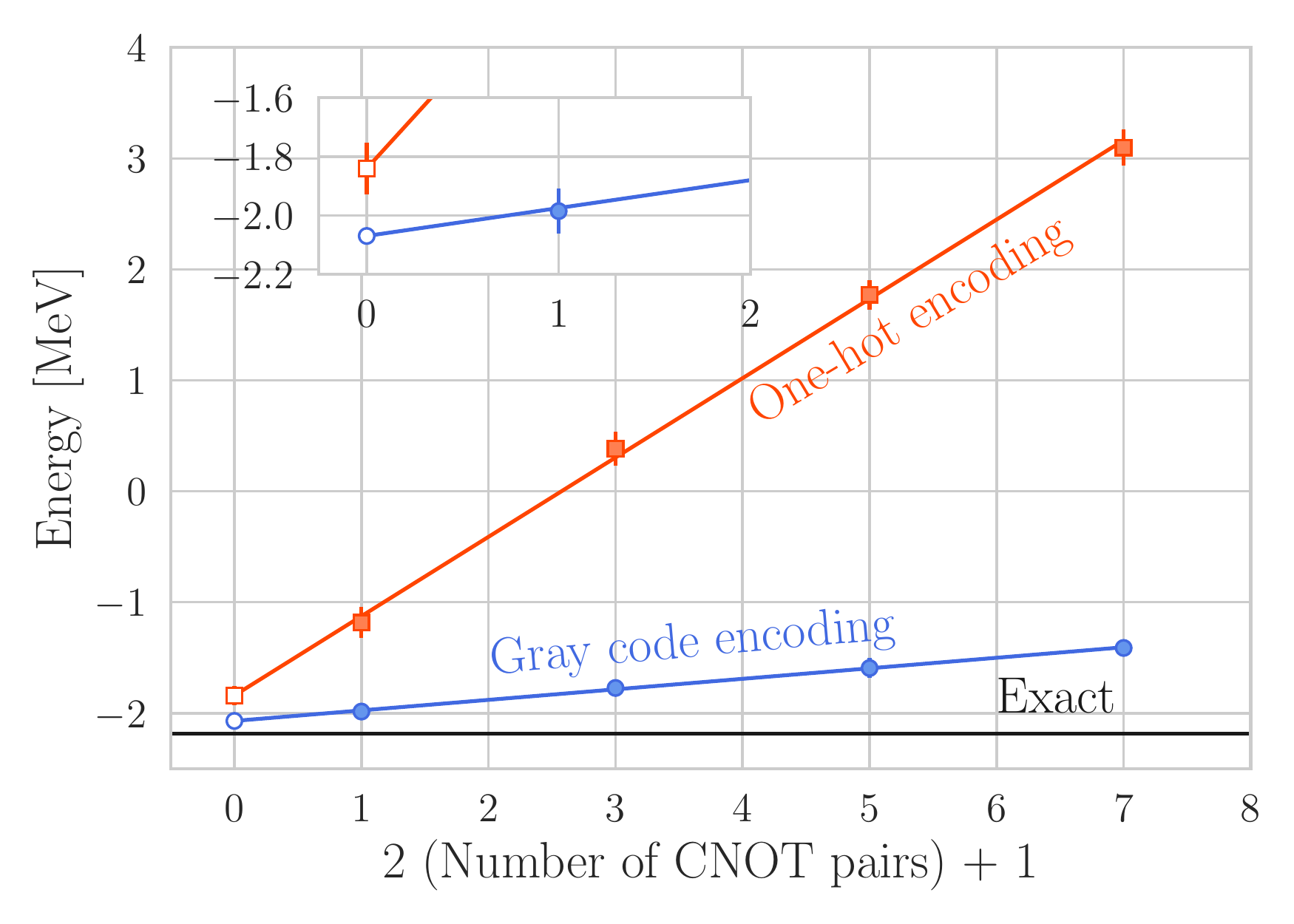}  
    \end{tabular}
    \caption{The energy extrapolation to the zero-noise limit.
    The filled symbols are the energies calculated with the IBM Q Vigo noise model, using the qubits \{2,1\} and \{2,1,3,4\} for the Gray code and one-hot encoding, respectively.
    Note that the variational parameters are optimized with classical computations.
    The solid lines indicate the linear fit using the filled symbol energies, and the extrapolated energies are shown as the unfilled symbols.
    The errors of input energies are evaluated as the standard deviation of the obtained distributions out of 100 trials with 10000 shots. 
    The errors of extrapolated energies are estimated as the root-mean-squared sum of the standard distribution of the energies and of the fitting procedure. The error bars on most of the Gray code points are too small to see. The errors on the one-hot points are between 0.13 and 0.16 while the Gray code errors are between 0.07 and 0.09.
    }
    \label{fig:zero_noise_extrapolation}
\end{figure}

To examine Eq.~(\ref{eq:extrap}), the extrapolations of the energies for both Gray code and one-hot encodings are demonstrated in ~\autoref{fig:zero_noise_extrapolation}.
To avoid considering uncertainties from the VQE process, the input energies with the IBM Q Vigo noise model (indicated by the solid symbols in the figure) are calculated with the optimal variational parameters, which are determined by classical calculations.
Since the slope is proportional to $\varepsilon_{0}$ which would be also roughly proportional to the number of CNOT gates, it is reasonable that 
 the slope for the Gray code encoding is smaller than that for the one-hot encoding.
Also, the linear extrapolation tends to worsen as $\varepsilon_{0}$ increases, i.e., the error of the linear fit in the one-hot case is larger than that of the Gray code case.
For the extrapolated energies, the Gray code energy seems slightly higher than the exact number, but still agrees within the error, while the one-hot energy is clearly off from the exact answer. 
One might think the discrepancy is due to the single-qubit gate error rate.
However, it was observed that there are no significant changes even if the single-qubit gate error is taken into account, using the unitary folding method discussed in \cite{UnitaryFolding}.
Further studies about zero-noise extrapolation would be needed to obtain a better estimation.

As zero-noise extrapolation does not quite yield the energy of the noiseless simulation, we combine extrapolation with the VQE process to try and improve these results.
To do this, the expectation value used at each VQE optimization step is evaluated by extrapolation using the CNOT pair insertion technique.
Employing the IBM Q Vigo noise model, the deuteron ground-state energies with 100 independent runs with 10000 shots are $E^{\rm (GC)}_{\rm g.s.}=-2.08 \pm 0.09$ MeV and $E^{\rm (OH)}_{\rm g.s.} = -1.89 \pm 0.14$ MeV with the Gray code and one-hot encodings, respectively.
Note that the uncertainties are estimated as the standard deviation of the distribution of 100 run results.

Since the extrapolation at each optimization step is done with the energies evaluated with the single 10000 shots calculations, which would be within the standard deviation of the distribution, it would be more reasonable to assign the uncertainty as the standard deviation than the standard error of the 100 independent runs.
Comparing to the exact energy $E^{\rm (exact)}_{\rm g.s.} = -2.14$ MeV, similarly to the energies evaluated at the optimal parameters (see Fig.~\ref{fig:zero_noise_extrapolation}), both encoding results provide higher energy than the exact, but the Gray code result agrees within the error.
Thus, the Gray code encoding yields better performance and enables us to estimate more accurate noiseless results.

\section{Application to Hamiltonian simulation}
\label{sec:hamsim}

This section presents an analysis of the Gray code encoding for simulating the time evolution of quantum systems. Simulating time evolution, or \emph{Hamiltonian simulation}, is one of the key applications of future large-scale quantum computers, but the resource requirements for nontrivial systems are beyond the capabilities of today's NISQ devices. Given a Hermitian Hamiltonian $H$, a unitary operation that performs evolution for time $t$ can be computed as
\begin{equation}
    U(t) = e^{-i H t}.
\end{equation}
The idea of Hamiltonian simulation using a quantum computer is to find (and then execute) a quantum circuit  $\tilde{U}(t)$ expressed in terms of elementary gates that approximates $U(t)$ well, i.e., $||\tilde{U}(t) - U(t)|| < \varepsilon$, where $\varepsilon$ is small and $||\cdot||$ is the spectral norm. Such a circuit is then typically used as a subroutine in a larger context, such as quantum phase estimation. A circuit can be found by first expressing $H$ as a linear combination of Paulis as in \autoref{eqn:ham_sum_paulis} \cite{Lloyd1996}:
\begin{equation}
    U(t) = e^{-i t \sum_j q_j Q_j} \approx  \prod_j e^{-i t q_j Q_j} + O(t^2).
\end{equation}
This expression is convenient because it involves sequential applications of unitaries of the form $e^{-itQ_j}$, which are straightforward to implement for a Pauli operator $Q_j$ \footnote{Consider a Pauli $Q$. To construct a circuit for $e^{-itQ}$, first note that since the Clifford group is the normalizer of the Pauli group, there exists a Clifford operation $C$ that diagonalizes $Q$, such that we can write $e^{-itQ}=e^{-it C^\dag Q_z C} = C^\dag e^{-itQ_z} C$ for some diagonal Pauli $Q_z$. $C$ can then be implemented using $H$ and $S$ gates. Then, as $Q_z$ is diagonal, $e^{-itQ_z}$ can be implemented using only Pauli $Z$ rotations.}. However an additional error operator term $O(t^2)$, the \emph{Trotter error}, arises due to the fact that for Pauli operators $Q_1$ and $Q_2$ that do not commute, $e^{Q_1 + Q_2} \neq e^{Q_1}e^{Q_2}$.  

The error can be decreased by \emph{Trotterization}. Intuitively, rather than evolving each term for time $t$, this process is divided up into $T$ repeated steps of time $t/T$, 
\begin{equation}
    U(t) = \left(\prod_j e^{-i q_j Q_j \frac{t}{T}} \right)^{T} + O(t^2/T),
    \label{eqn:trotterized_ham}
\end{equation}
$T$ is known as the \emph{Trotter number}, or number of Trotter steps. There also exist higher-order versions of this formula \cite{Suzuki1976}, though only the first-order expansion as written in \autoref{eqn:trotterized_ham} is used in this investigation.

\begin{figure}[ht]
    \centering
    \includegraphics[width=0.5\textwidth]{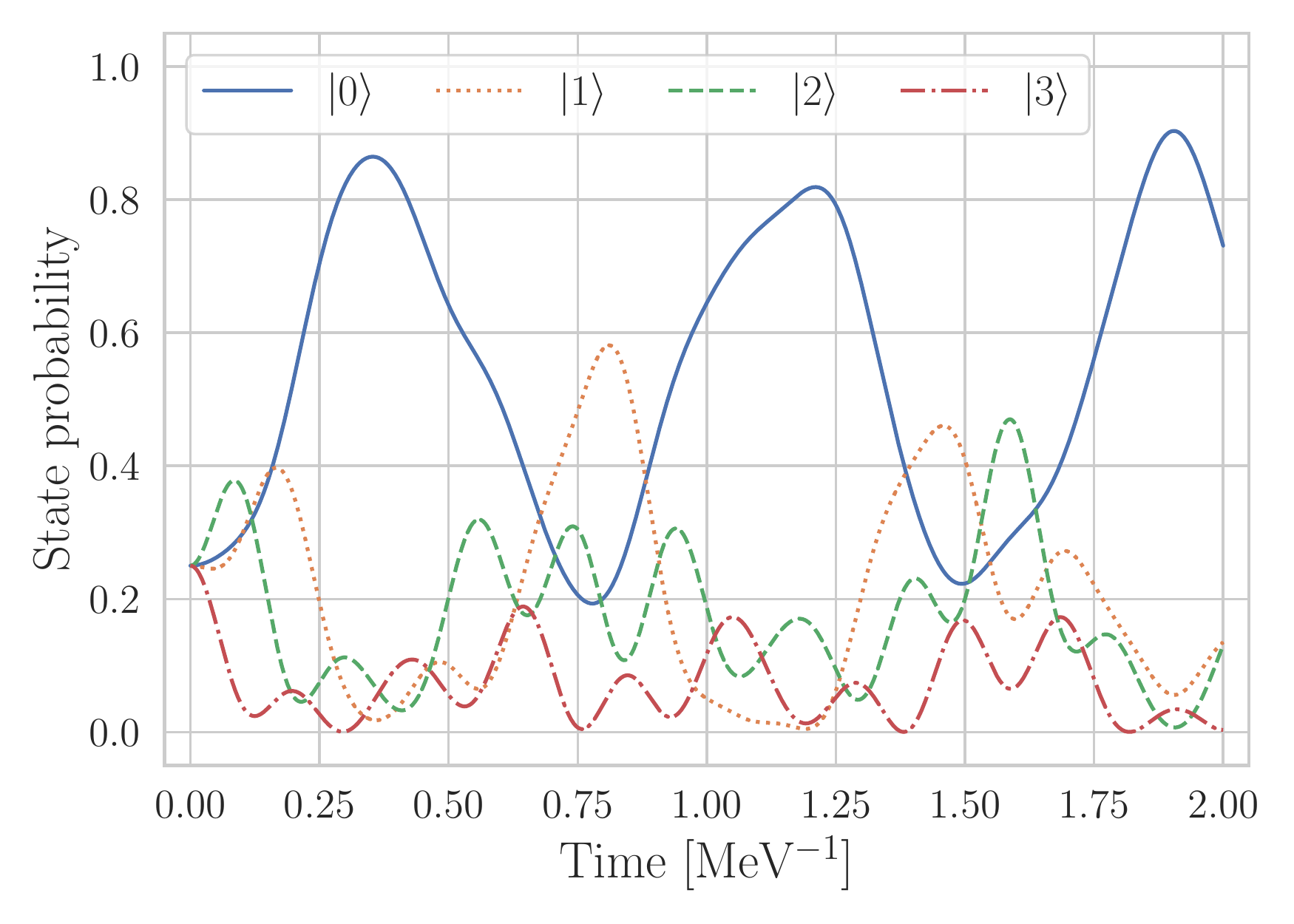}
    \caption{Evolution over time of the basis-state probabilities under $N=4$ deuteron Hamiltonian [$U(t)=\exp(-itH_4)$].  In later figures the simulation at $t=1$ is investigated in detail.}
    \label{fig:extended_time_evolution}
\end{figure}

To compare the two encodings, the $N=4$ case is explored. The evolution of the probability distribution of basis-state measurement outcomes is shown in \autoref{fig:extended_time_evolution} for reference. In subsequent analysis, the evolution time $t$ is set arbitrarily to $t = 1$. To perform the evolution, the system is first initialized in the uniform superposition of its relevant basis states (occupation basis for the one-hot case, and full computational basis for the Gray code case), followed by Hamiltonian evolution for a number of Trotter steps, which varies from 1 to 100.

The evolution circuits are generated using the \texttt{evolve} functionality of Qiskit Aqua's \texttt{WeightedPauliOperator} class. Optimization of these circuits using the Qiskit transpiler was performed. The transpiler contains four preset levels of optimization;  while all four levels were investigated, results of only the two highest levels are reported here. Going from level 0 to level 1 yielded a reduction of single qubit gates by roughly 50\% for both encodings; further optimization to level 2 did not result in any improvements over level 1   in either case, so level 2 is used here as the representative level. A substantial improvement was observed with level 3 for the Gray code case, so this level must also be considered.

\begin{figure}[ht]
 \includegraphics[width=0.95\textwidth]{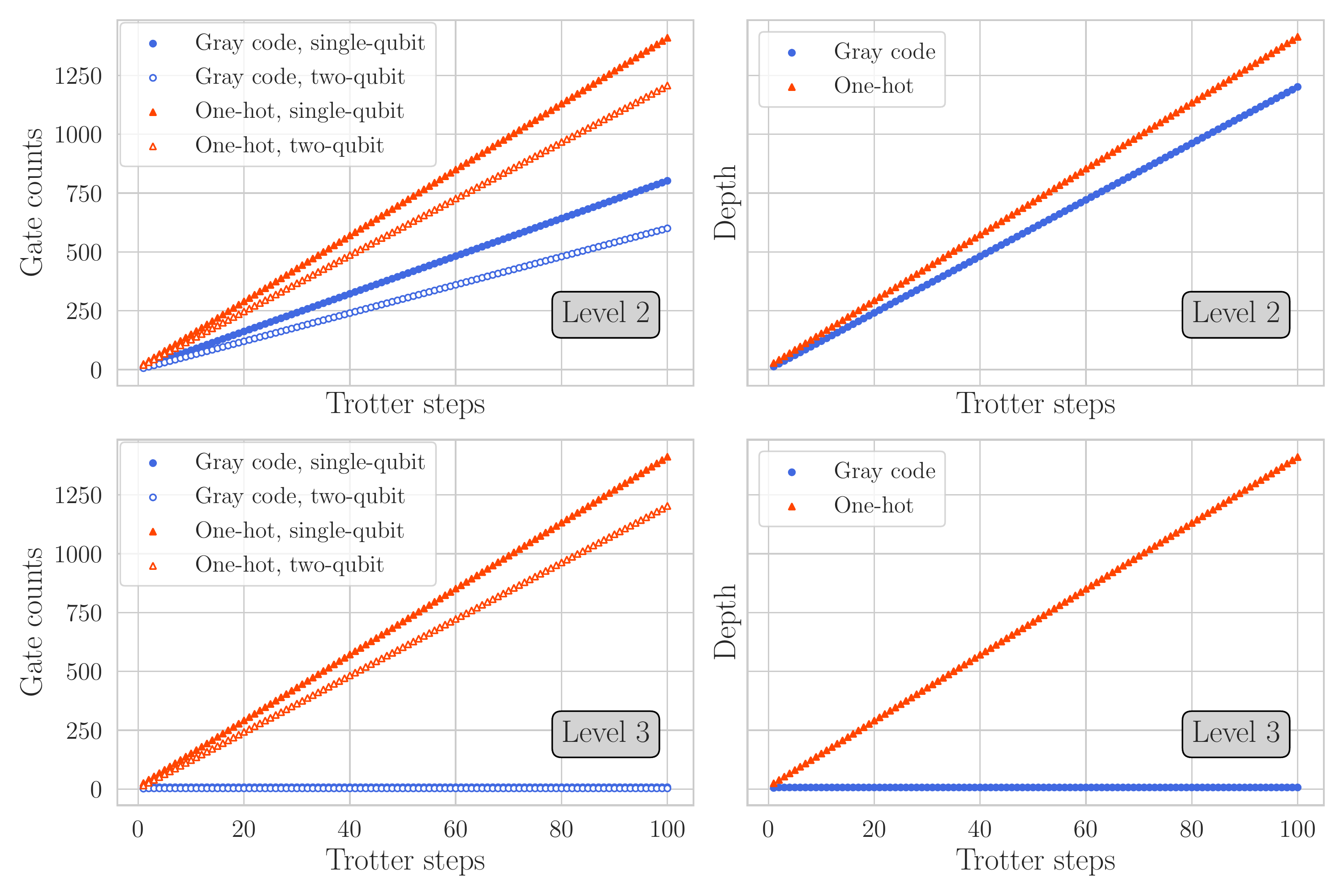}
\caption{Resource requirements for Hamiltonian evolution circuits of the deuteron Hamiltonian for $N=4$ at two different levels of circuit optimization.  In the level 2 optimization, the CNOT count of the Gray code encoding circuit is about $50\%$ that of the one-hot version. For single-qubit gates, the Gray code uses about $60\%$ of the amount. The depth of the Gray code circuit is roughly $85\%$ that of the equivalent one-hot circuit. At optimization level 3 two-qubit unitary resynthesis leads to Gray code circuits of fixed size and depth, independent of the number of Trotter steps.}
\label{fig:time_evolution_circuit_resources}
\end{figure}

 The circuit resources are plotted in \autoref{fig:time_evolution_circuit_resources}, where it can be observed that the Gray code encoding circuits require roughly half the amount of CNOT gates, nearly 60\% of the single-qubit gates, and can be executed in 85\% of the depth of the one-hot encoding circuits. These results  demonstrate the extent to which Hamiltonian evolution could be performed with fewer resources when using the Gray code encoding, though we note that for both encodings the gate count and depth are far beyond what can be done with a NISQ-era machine, as will be demonstrated in the  numerical results of \autoref{fig:fig12-hamsim-noise}. 

\begin{figure}[ht]
  \centering
  \includegraphics[width=0.8\textwidth]{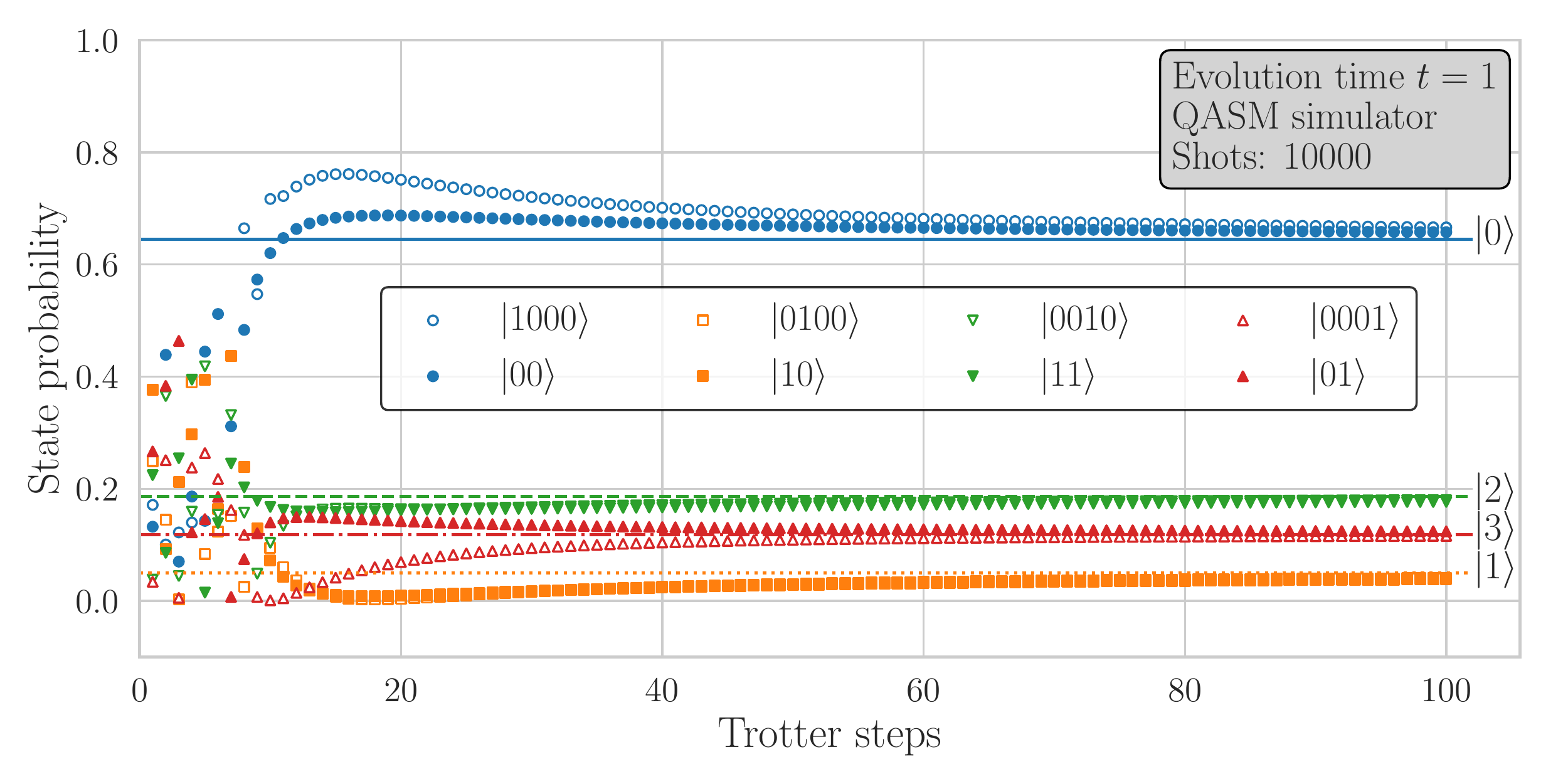}
\caption{Comparison of state probabilities at fixed time $t=1$ obtained using standard Trotter decomposition for Hamilton simulation with the $N=4$ deuteron Hamiltonian for Gray code (filled) and one-hot (unfilled) encodings. Results obtained using 10000 QASM shots for each amount of Trotter steps. The lines show the true value for each state $\ket{n}$ ($n=0,1,2,3$) computed analytically by exponentiating the Hamiltonian and applying it to the uniform superposition.}
\label{fig:qasm_time_evolution}
\end{figure}

As a baseline, the evolution is first analyzed in an ideal setting, and the probability distribution of each basis state after evolution was estimated using the QASM simulator with 10000 shots. An example of this is plotted in \autoref{fig:qasm_time_evolution}, where as expected the probability distributions approach the true values as the number of Trotter steps is increased.

\begin{figure}[ht]
  \centering
  \includegraphics[width=0.5\textwidth]{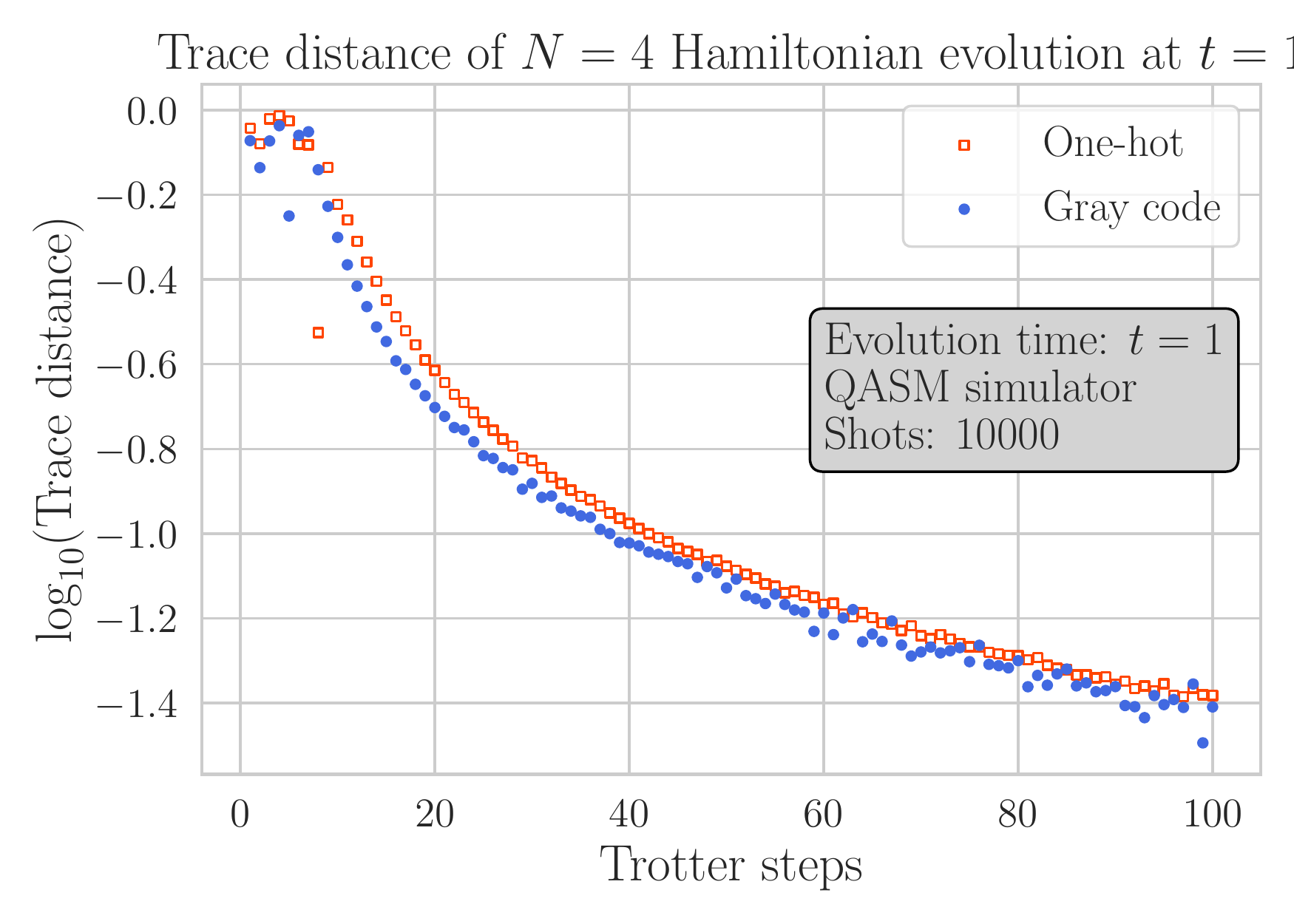}
\caption{Comparison of trace distance (\autoref{eqn:trace-distance}) for both encodings at time $t=1$.  Trace distances are calculated between the density matrix computed using state tomography on QASM runs with 10000 shots for the given amount of Trotter steps, and the density matrix computed analytically (no simulated hardware noise is present). The trace distances are quite similar between the two encodings. As they quickly approach 0 in both cases, the $\log_{10}$ is plotted to emphasize differences.}
\label{fig:fig10a-hamsim-qasm-tracedist}
\end{figure}

To quantify the quality of the evolution circuit, the trace distance is taken between the output state after evolution and the ideal output state as computed directly from $e^{-iHt} \ket{\psi}$ for the uniform superposition $\ket{\psi}$. For two quantum states represented by density matrices $\rho$ and $\rho^\prime$, the trace distance is defined as \cite{NielsenChuang}
\begin{equation}
 D(\rho, \rho^\prime) = \frac{1}{2} \hbox{Tr}|\rho - \rho^\prime|,
 \label{eqn:trace-distance}
\end{equation}
where the norm $|A| = \sqrt{A^\dag A}$ (thus smaller trace distance is better). A density matrix for the output state from the QASM simulations is estimated using state tomography with Qiskit's Ignis library. In \autoref{fig:fig10a-hamsim-qasm-tracedist} the trace distances are plotted. It can be seen that while the Gray code encoding fares slightly better, the trace distances are comparable, and as expected both decrease as the number of Trotter steps increases.

As with the VQE, the situation of greater interest is when hardware noise is present. The same simulations were repeated using the Vigo device noise model (including measurement error mitigation), and the resultant trace distances are plotted in \autoref{fig:fig12-hamsim-noise}(a) for level 2 optimization, and \autoref{fig:fig12-hamsim-noise}(b) for level 3 optimization.

\begin{figure}[ht]
  \centering
  \includegraphics[width=0.5\textwidth]{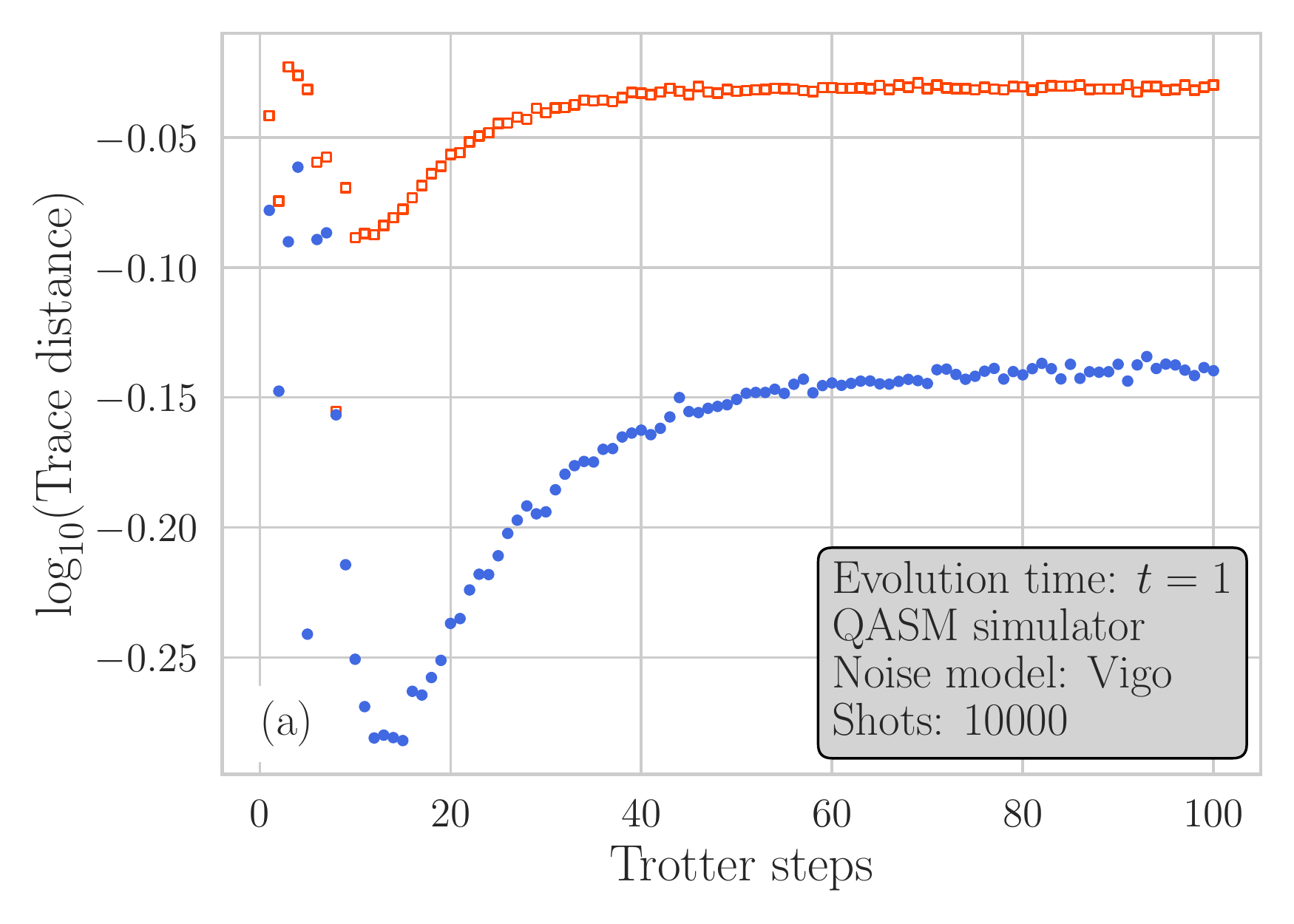}
  \includegraphics[width=0.5\textwidth]{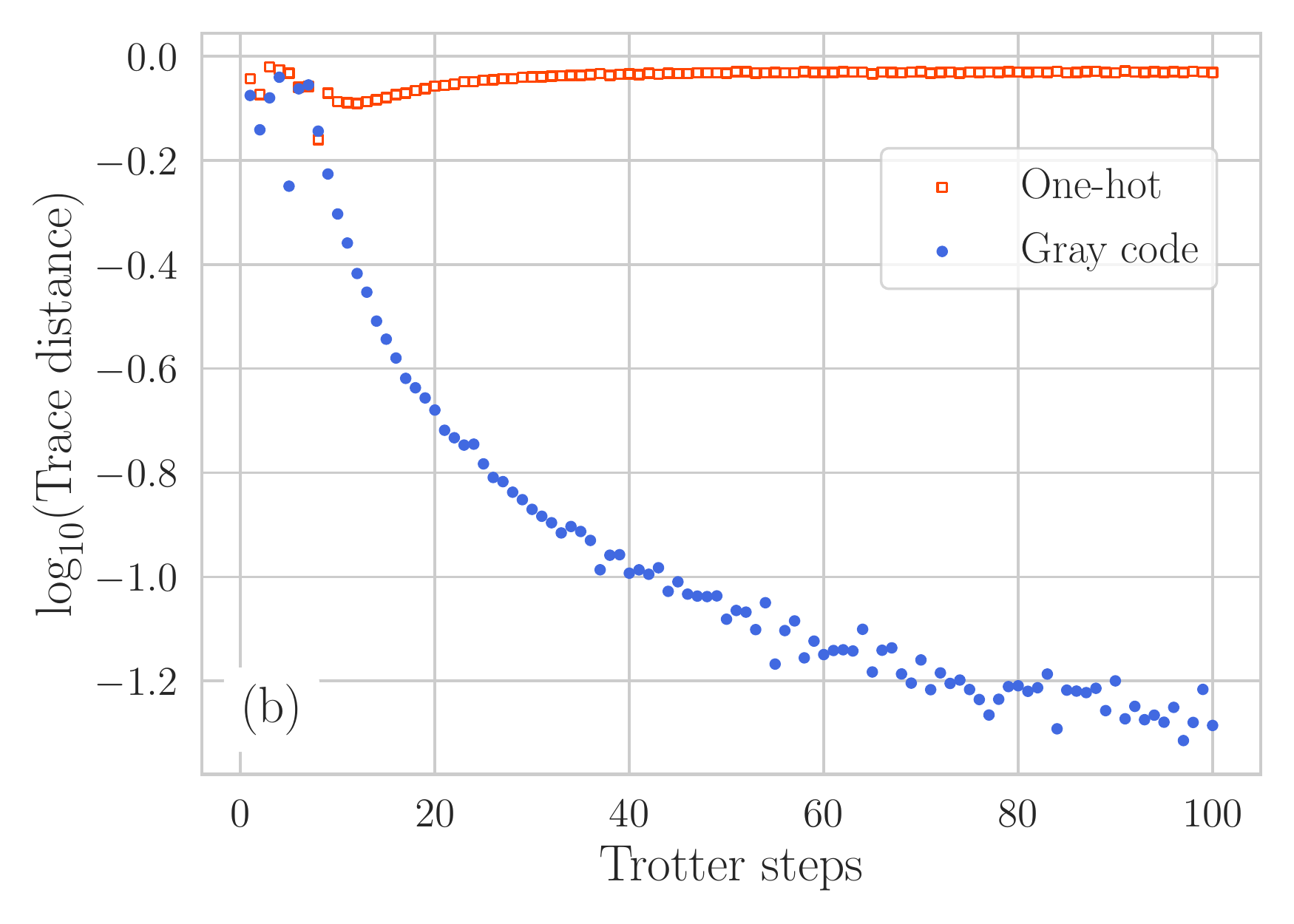}
    \caption{Comparison of trace distances (\autoref{eqn:trace-distance}) between encodings at time $t=1$ with Vigo noise model.  Trace distances are calculated between the density matrix computed using state tomography with measurement error mitigation applied on QASM runs with 10000 shots for given amount of Trotter steps, and the density matrix computed analytically.    For the Gray code encoding, the initial qubit placement was \{2, 1\}. For the one-hot encoding, it is \{2, 1, 3, 4\}, which is a perfect assignment with no additional SWAP gates added. 
    (a) Level 2 circuit optimization. The value of the trace distances at the plateau correspond roughly to the trace distance of the expected output state with the maximally mixed state, indicating that the system has fully decohered.
    (b) Level 3 circuit optimization. The additional optimization step does not change the one-hot encoding results. However, level 3 optimization includes resynthesis of two-qubit unitaries, which effectively collapses the entire circuit down to a size independent of the number of Trotter steps, leading to results that parallel the ideal case for the Gray code encoding.
    }
    \label{fig:fig12-hamsim-noise}

\end{figure}

In the level 2 case, both encodings see the trace distances initially improve as the number of Trotter steps increases, which is expected as using more Trotter steps produces more accurate simulations. However in both cases, this improvement eventually ceases and the trace distance begins to increase and then plateau. For both encodings, the value at the plateau corresponds roughly to the trace distance between the true expected output (computed analytically), and the maximally mixed state, indicating total decoherence of the system. The turning point occurs at a larger number of Trotter steps for the Gray code encoding, around 15 Trotter steps rather than 10 Trotter steps. This aligns directly with the level 2 depth plot of \autoref{fig:time_evolution_circuit_resources} in which it can be seen that the depth of the Gray code circuit at 15 Trotter steps is close to that of the one-hot circuits at 10. Past this point, both encodings are limited by the hardware noise.

The level 3 case yields starkly different results. Level 3 optimization with Qiskit's transpiler applies two-qubit unitary resynthesis to the circuits. As the Gray code circuits consist of only 2 qubits, they are always resynthesized down to a small sequence of one- and two-qubit gates. This is the reason for the higher-quality results presented in \autoref{fig:fig12-hamsim-noise}(b). While this is a special case afforded to us by virtue of the number of qubits, it highlights the value of exploring the trade-offs between different encodings and end-to-end treatment of the problem with full optimization, since in some cases substantial improvements may be possible. 

There are a number of additional aspects of Hamiltonian simulation for which the two encodings should be compared in future work. A key one is to quantify the Trotter error, and investigate the effect of the partitioning of the Paulis into commuting sets. For the deuteron Hamiltonian with $N$ states, recall that the Gray code encoding yields $N + 1$ sets of commuting Paulis, while one-hot yields 3.  Furthermore, the order in which to perform the product terms of \autoref{eqn:trotterized_ham} is a subject of active investigation \cite{Tranter2019}. Studying different Trotter decomposition formulas, or even different Hamiltonian simulation schemes such as qubitization \cite{Low_2019}, might also reveal interesting differences.

\section{Conclusions and future work}
\label{sec:conclusions}

A mapping that orders the computational basis states in a Gray code yields a number of advantages in the context of the VQE. It requires exponentially fewer qubits compared to the same problem solved using the one-hot encoding, and can use smaller, hardware-efficient variational ansatze that require fewer CNOT gates for problems solvable on NISQ-era machines.  It also suggests a natural partitioning into $N + 1$ sets of commuting operators where only one qubit's basis must be rotated during measurement. While the number of measurements is larger and also increases with system size, this method may nevertheless be beneficial in the near term due to the trade-off with number of qubits and gate counts, as the variance in energies produced by the VQE is reduced (most notably in the presence of simulated hardware noise). Similar advantages are observed for performing Hamiltonian simulation, where using the Gray code encoding enables us to perform evolution using fewer resources than the one-hot encoding.

In order to demonstrate any long-term advantages, the Gray code encoding must be extended and adapted to non-trivial situations. There are a number of future directions to be investigated. First, testing must be done on actual hardware to analyze how additional limitations affect the solution quality.  The deuteron is fully solvable using present-day classical methods, and so an immediate next step is to extend and test the method in a  multiparticle scenario. This could be done by simply concatenating a set of registers, one for each particle, with each expressed independently in a Gray code. However, care must be taken to ensure antisymmetrization of the basis is satisfied.  

Another avenue is to extend the Gray code encoding to calculating the ground state of the deuteron using a full \emph{ab initio} chiral interaction~\cite{MACHLEIDT2011,Epelbaum2015,EMN2017}.  The inclusion of higher-order chiral terms in the interaction will result in more complicated ladder operators, which will likely remove the useful property of having only one $X$ per commuting Pauli set, making simultaneous measurement more complex. Especially in these more complex cases, the tradeoffs with the one-hot case must be studied.

The cyclic nature of the Gray code naturally suggests another application: working on a periodic lattice. Similar work that focuses on lattice methods indexes the lattice sites using qubits in binary order (for example \cite{Roggero2019}). This yields more complicated transformations when moving across the boundaries of the lattice, as one has to make the transition from $\ket{1\cdots 1}$ back to $\ket{0\cdots0}$. Indexing using a Gray code, which can be done over both dimensions of a 2D lattice, will simplify these transitions.

Further analysis of time evolution is also necessary, in particular the effect of the encoding on the amount of Trotter error since there are more commuting sets of Paulis. Time evolution should also be extended to the multiparticle case, and investigated for different decompositions such as higher-order Trotter formulae, and Trotter-Suzuki decompositions. While the gate counts for this are far beyond the capabilities of NISQ hardware, it may still enable us to simulate larger systems sooner by making better use of available resources.


\acknowledgments

We thank Martin Savage and Alessandro Roggero for valuable discussions. This work was in part supported from NSERC grants No. SAPIN-2016-00033 and No. PGSD3-535536-2019. TRIUMF receives federal funding via a contribution agreement with the National Research Council of Canada. Computations were performed on the Oak Cluster at TRIUMF managed by Advanced Research Computing (ARC) at the University of British Columbia. We acknowledge the use of IBM Quantum services for this work. The views expressed are those of the authors, and do not reflect the official policy or position of IBM or the IBM Quantum team.

\appendix

\section{Mapping using a standard basis ordering}
\label{appendix:why-not-regular-order}

As the Gray code is a reordering of the computational basis states, one might wonder what happens if the basis states were simply ordered in increasing binary value. While this can certainly be done, the Gray code encoding simplifies the measurement process in the VQE and thus reduces the number of gates that must be applied. As an example, consider the $N=4$ case, whose operators are detailed in \autoref{tab:standard_2qubit}.

{\begin{table}[ht]    
    \caption{Mapping from operators of the deuteron Hamiltonian  \eqref{eq:basic_ham} acting on the harmonic oscillator (HO) basis with $N=4$ to a two-qubit system using the standard computational basis ordering.
    \label{tab:standard_2qubit}
    }
\begin{subtable}{\linewidth}
    \centering    
    \caption{Number-operator terms}
    \begin{tabular}{|c|c|c|}
     \hline
          HO states & Qubit operator & Qubit states \\ \hline
          $\ket{0}\bra{0}$ & $P^{(0)}_0 P^{(0)}_1$ & $\ket{00}\bra{00}$\\ \hline
          $\ket{1}\bra{1}$ & $P^{(1)}_0 P^{(0)}_1$ & $\ket{10}\bra{10}$\\ \hline
          $\ket{2}\bra{2}$ & $P^{(0)}_0 P^{(1)}_1$ & $\ket{01}\bra{01}$\\ \hline
          $\ket{3}\bra{3}$ & $P^{(1)}_0  P^{(1)}_1$ & $\ket{11}\bra{11}$\\ \hline
    \end{tabular}

    \label{tab:standard_number_terms_2qubit}
\end{subtable}

\begin{subtable}{\linewidth}
\vspace{.2cm}
    \centering
        \caption{Ladder-operator terms.}
    \begin{tabular}{|c|c|c|}
     \hline
          HO states & Qubit operator & Qubit states \\ \hline
          $\ket{0}\bra{1}$ & $ X_0P^{(0)}_1$ & $\ket{00}\bra{10}$\\ \hline
          $\ket{1}\bra{2}$ & $X_0 X_1$ & $\ket{10}\bra{01}$\\ \hline
          $\ket{2}\bra{3}$ & $ X_0 P^{(1)}_1$ & $\ket{01}\bra{11}$\\ \hline
    \end{tabular}

    \label{tab:standard_ladder_terms_2qubit}
    \end{subtable}

\end{table}
}

The Paulis that will be present in the Hamiltonian are:
\begin{equation}
    \unit, \enskip Z_0, \enskip Z_1, \enskip Z_0 Z_1, \enskip X_0,\enskip X_0 Z_1,\enskip X_0 X_1
\end{equation}
These can be partitioned into 3 commuting sets, $S_Z = \{Z_0, Z_1, Z_0 Z_1 \}$, $S_X = \{X_0, X_0 X_1 \}$, and $S_{XZ} = \{ X_0 Z_1 \}$. Simultaneous measurements of $S_Z$ are done simply with the computational basis. For $S_X$ the basis of \emph{all} qubits must be rotated by a Hadamard to perform the measurement. For $S_{XZ}$ only the basis of the first qubit must be rotated. Using this standard ordering on any $N$-qubit system will always produce a term in the Hamiltonian containing $X_0 \cdots X_{N-1}$, which will require rotation of all $N$ qubits prior to measurement (and similarly measurements that must rotate $N-1$ qubits, $N-2$, and so on). The total number of basis rotations for a full set of measurements is thus $N(N+1)/2 = O(N^2)$, which is higher than the $N$ required when using the Gray code encoding on the same system.

\section{Gray code encoding tables}
\label{appendix:encoding-tables}

This appendix contains the tables of number and ladder operators of the Gray code encoding for $N=4$ (\autoref{tab:gc_2qubit}) and $N=8$ (\autoref{tab:gc_3qubit}).

{\begin{table}[h!]

    \caption{Mapping from operators of the deuteron Hamiltonian  \eqref{eq:basic_ham} acting on the harmonic oscillator (HO) basis with $N=4$ to a two-qubit system using the Gray code encoding.
    \label{tab:gc_2qubit}
    }
\begin{subtable}{\linewidth}
    \centering
    \caption{Number-operator terms.}
    \begin{tabular}{|c|c|c|}
     \hline
          HO states & Qubit operator & Qubit states \\ \hline
          $\ket{0}\bra{0}$ & $P^{(0)}_0 P^{(0)}_1$ & $\ket{00}\bra{00}$\\ \hline
          $\ket{1}\bra{1}$ & $P^{(1)}_0 P^{(0)}_1$ & $\ket{10}\bra{10}$\\ \hline
          $\ket{2}\bra{2}$ & $P^{(1)}_0 P^{(1)}_1$ & $\ket{11}\bra{11}$\\ \hline
          $\ket{3}\bra{3}$ & $P^{(0)}_0  P^{(1)}_1$ & $\ket{01}\bra{01}$\\ \hline
    \end{tabular}
    \label{tab:gc_number_terms_2qubit}
\end{subtable}

\begin{subtable}{\linewidth}
\vspace{.2cm}
    \caption{Ladder-operator terms.}
    \centering
    \begin{tabular}{|c|c|c|}
     \hline
          HO states & Qubit operator & Qubit states \\ \hline
          $\ket{0}\bra{1}$ & $ X_0P^{(0)}_1$ & $\ket{00}\bra{10}$\\ \hline
          $\ket{1}\bra{2}$ & $P^{(1)}_0X_1$ & $\ket{10}\bra{11}$\\ \hline
          $\ket{2}\bra{3}$ & $ X_0P^{(1)}_1$ & $\ket{11}\bra{01}$\\ \hline
    \end{tabular}
    \label{tab:gc_ladder_terms_2qubit}
    \end{subtable}
    
\end{table}
}

\begin{table}[h!]
    \caption{Mapping from operators of the deuteron Hamiltonian  \eqref{eq:basic_ham} acting on the harmonic oscillator (HO) basis with $N=8$ to a three-qubit system using the Gray code encoding.
    }
\begin{subtable}{\linewidth}
    \centering
        \caption{Number-operator terms.}
    \begin{tabular}{|c|c|c|}
     \hline
          HO states & Qubit operator & Qubit states \\ \hline
          $\ket{0}\bra{0}$ & $P^{(0)}_0  P^{(0)}_1  P^{(0)}_2$ & $\ket{000}\bra{000}$\\ \hline
          $\ket{1}\bra{1}$ & $P^{(1)}_0  P^{(0)}_1 P^{(0)}_2$ & $\ket{100}\bra{100}$\\ \hline
          $\ket{2}\bra{2}$ & $P^{(1)}_0 P^{(1)}_1 P^{(0)}_2$ & $\ket{110}\bra{110}$\\ \hline
          $\ket{3}\bra{3}$ & $P^{(0)}_0  P^{(1)}_1 P^{(0)}_2$ & $\ket{010}\bra{010}$\\ \hline
          $\ket{4}\bra{4}$ & $P^{(0)}_0  P^{(1)}_1 P^{(1)}_2$ & $\ket{011}\bra{011}$\\ \hline
          $\ket{5}\bra{5}$ & $P^{(1)}_0  P^{(1)}_1 P^{(1)}_2$ & $\ket{111}\bra{111}$\\ \hline
          $\ket{6}\bra{6}$ & $P^{(1)}_0  P^{(0)}_1 P^{(1)}_2$ & $\ket{101}\bra{101}$\\ \hline
          $\ket{7}\bra{7}$ & $P^{(0)}_0  P^{(0)}_1 P^{(1)}_2$ & $\ket{001}\bra{001}$\\ \hline
    \end{tabular}

        \label{tab:gc_number_terms_3qubit}
\end{subtable}
\begin{subtable}{\linewidth}
\vspace{.2cm}
    \centering
        \caption{Ladder-operator terms.}
    \begin{tabular}{|c|c|c|}
     \hline
          HO states & Qubit operator & Qubit states \\ \hline
         $\ket{0}\bra{1}$ & $X_0  P^{(0)}_1  P^{(0)}_2$ & $\ket{000}\bra{100}$\\ \hline
          $\ket{1}\bra{2}$ & $P^{(1)}_0  X_1 P^{(0)}_2$ & $\ket{100}\bra{110}$\\ \hline
          $\ket{2}\bra{3}$ & $X_0 P^{(1)}_1 P^{(0)}_2$ & $\ket{110}\bra{010}$\\ \hline
          $\ket{3}\bra{4}$ & $P^{(0)}_0  P^{(1)}_1 X_2$ & $\ket{010}\bra{011}$\\ \hline
         $\ket{4}\bra{5}$ & $X_0  P^{(1)}_1 P^{(1)}_2$ & $\ket{011}\bra{111}$\\ \hline
         $\ket{5}\bra{6}$ & $P^{(1)}_0  X_1 P^{(1)}_2$ & $\ket{111}\bra{101}$\\ \hline
          $\ket{6}\bra{7}$ & $X_0  P^{(0)}_1 P^{(1)}_2$ & $\ket{101}\bra{001}$\\ \hline
    \end{tabular}

    \label{tab:gc_ladder_terms_3qubit}
    \end{subtable}

    \label{tab:gc_3qubit}
\end{table}

\section{Additional Noise Models}
\label{appendix:additional-noise-models}

\begin{figure}
    \begin{tabular}{cc}
        \includegraphics[width=0.45\textwidth]{fig5-ibmq_vigo_graph.pdf}  &
        \includegraphics[width=0.45\textwidth]{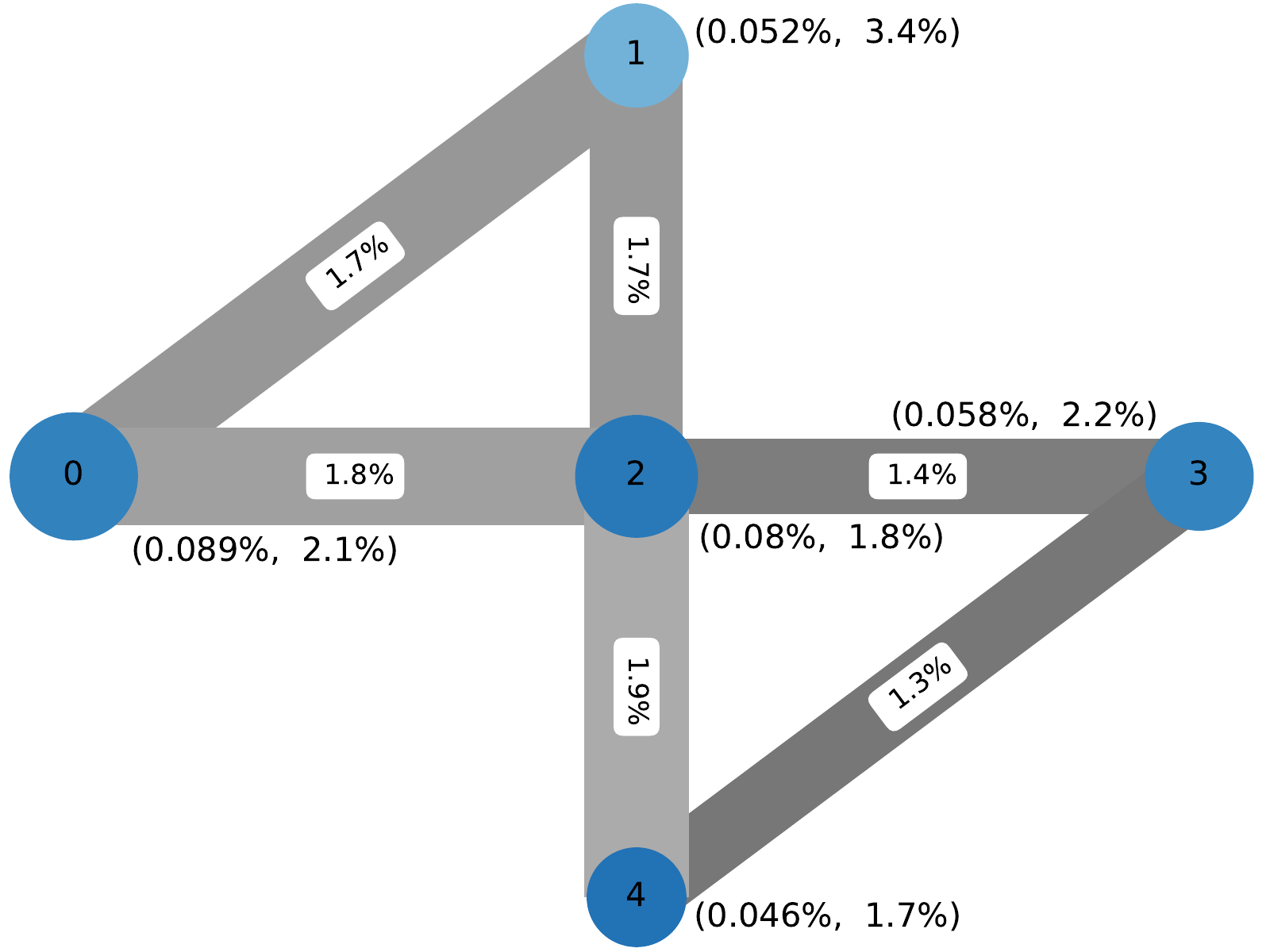} \\
    \end{tabular}
    \caption{The simulated hardware for IBM Q Vigo (left) and Yorktown (right). Each node of the graph corresponds to a physical qubit. The pair of values in the label correspond to the single-qubit gate error rate and measurement error rate, respectively. The value on the edges corresponds to the two-qubit gate error rate. Lighter color, larger node size and larger edge width correspond to higher error rates.}
    \label{fig:hardware_graphs_appendix}
\end{figure}

\begin{figure}
    \begin{tabular}{cc}
        \includegraphics[width=.5\textwidth]{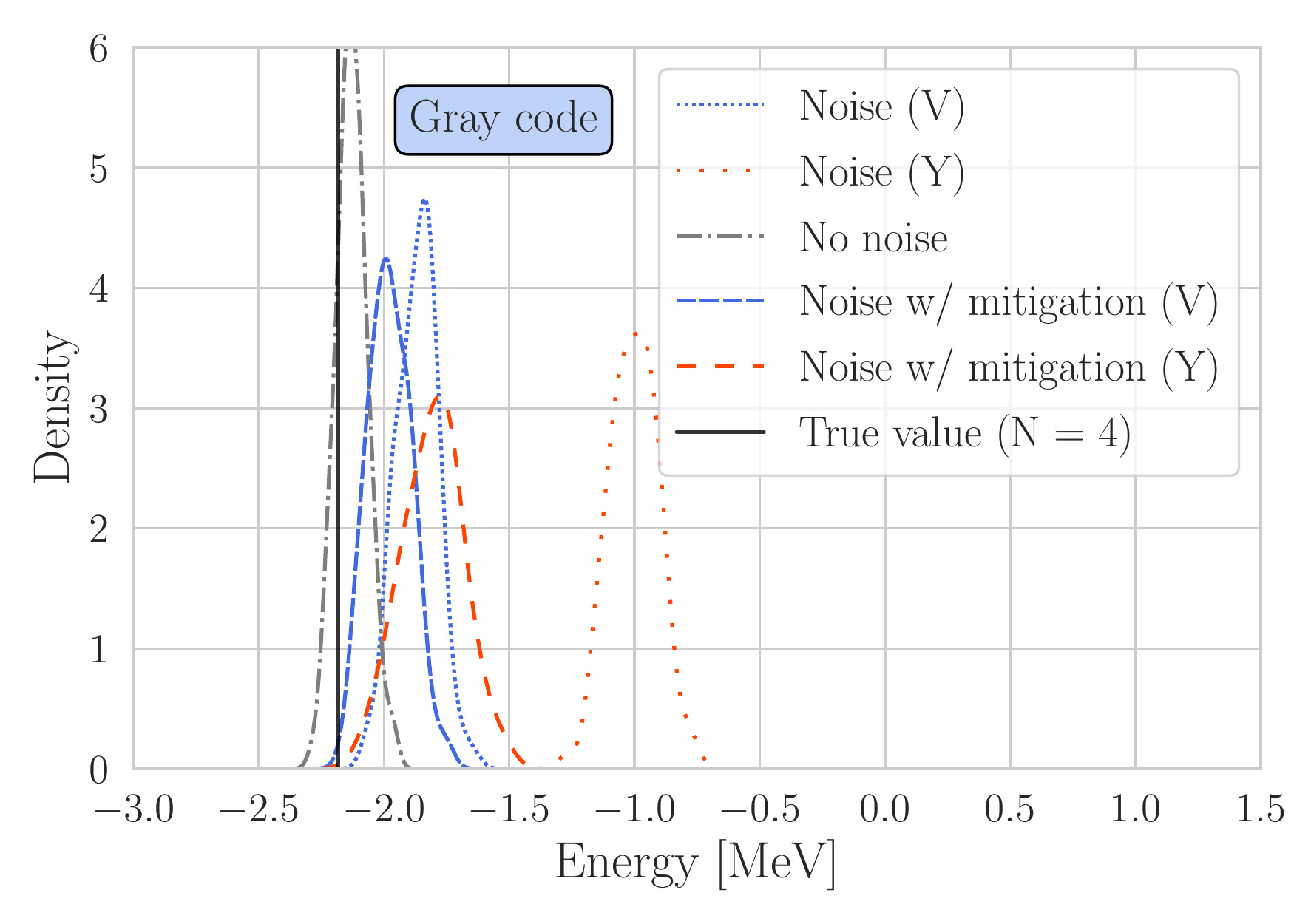} &
        \includegraphics[width=.5\textwidth]{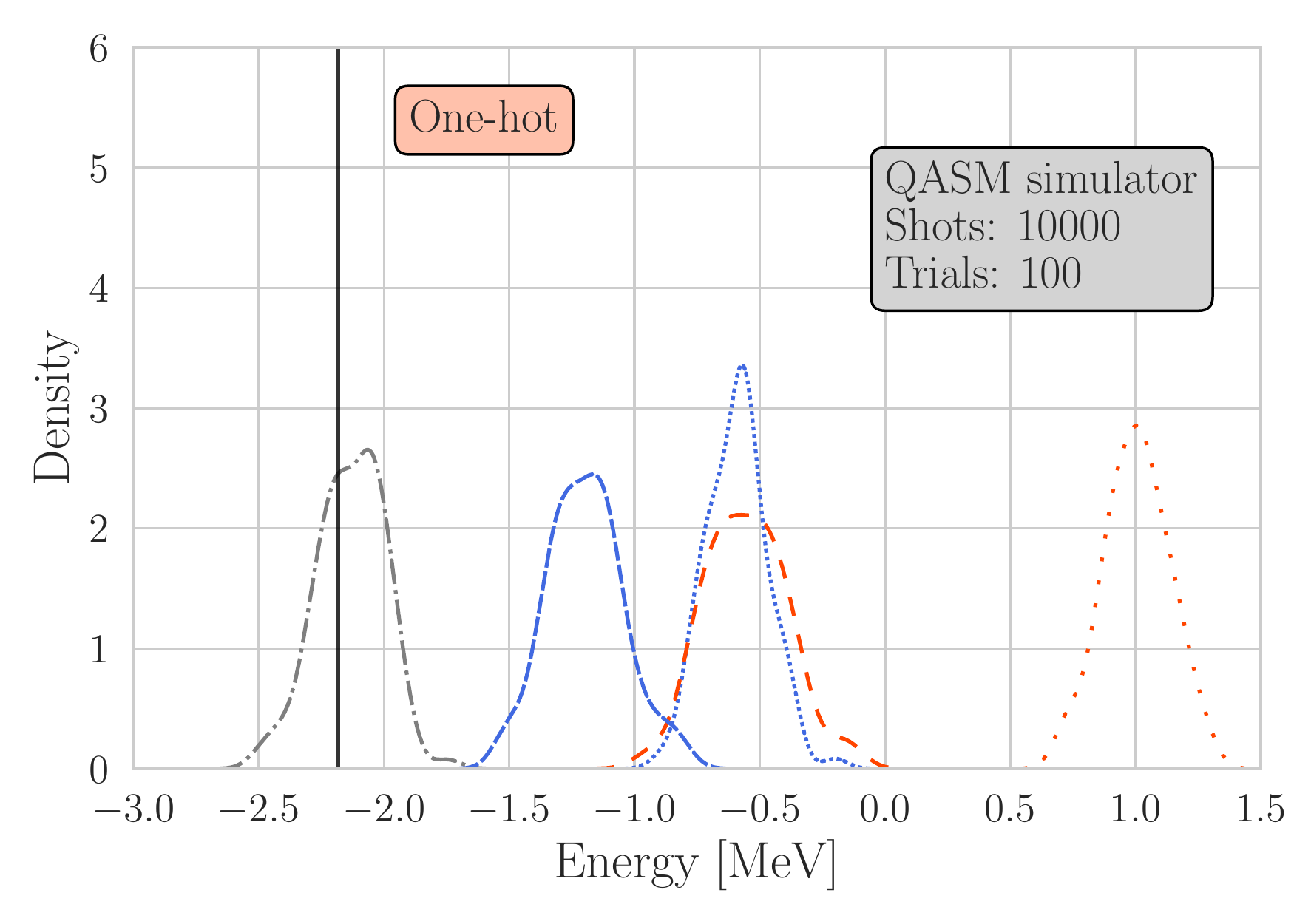} \\
    \end{tabular}
    \caption{The distribution of VQE energies for QASM simulations of the deuteron Hamiltonian for $N=4$  using the Vigo and Yorktown device noise models. In the legend, Vigo is denoted by (V) and Yorktown by (Y). The Vigo results are identical to those of Figure \ref{fig:noise_comparison}. For the Yorktown simulations, the Gray code maps the virtual qubits \{0,1\} to the ``physical'' qubits \{2,1\}, and the one-hot simulation maps \{0,1,2,3\} to \{0,1,2,3\}.}
    \label{fig:noise_comparison_appendix}
\end{figure}

The hardware errors experienced by NISQ-era physical quantum computers depend on parameters which can change over time.
Noise models based on IBM devices are recalibrated daily, therefore the results obtained using these models cannot be expected to be replicated exactly.
However, the improvement of VQE results achieved using the Gray code should be consistent.
In this section, the same simulations as in Fig. \ref{fig:noise_comparison} are carried out on a second noise model of a different device (IBM Q Yorktown \footnote{\emph{ibmq\_5\_yorktown - ibmqx2} v2.1.0, IBM Quantum team. Retrieved from https://quantum-computing.ibm.com (2020)}) to verify the results of Section 4.
Figure \ref{fig:hardware_graphs_appendix} compares the hardware graphs of Vigo and Yorktown; the latter has significantly higher error rates.
Figure \ref{fig:noise_comparison_appendix} compares the VQE results with different encodings for both devices.
The greater noise in Yorktown results in a shift of the energies relative to Vigo, but in both cases the Gray code encoding performs better than the one-hot encoding.

\begin{figure}
    \begin{tabular}{c}
        \includegraphics[width=0.5\textwidth]{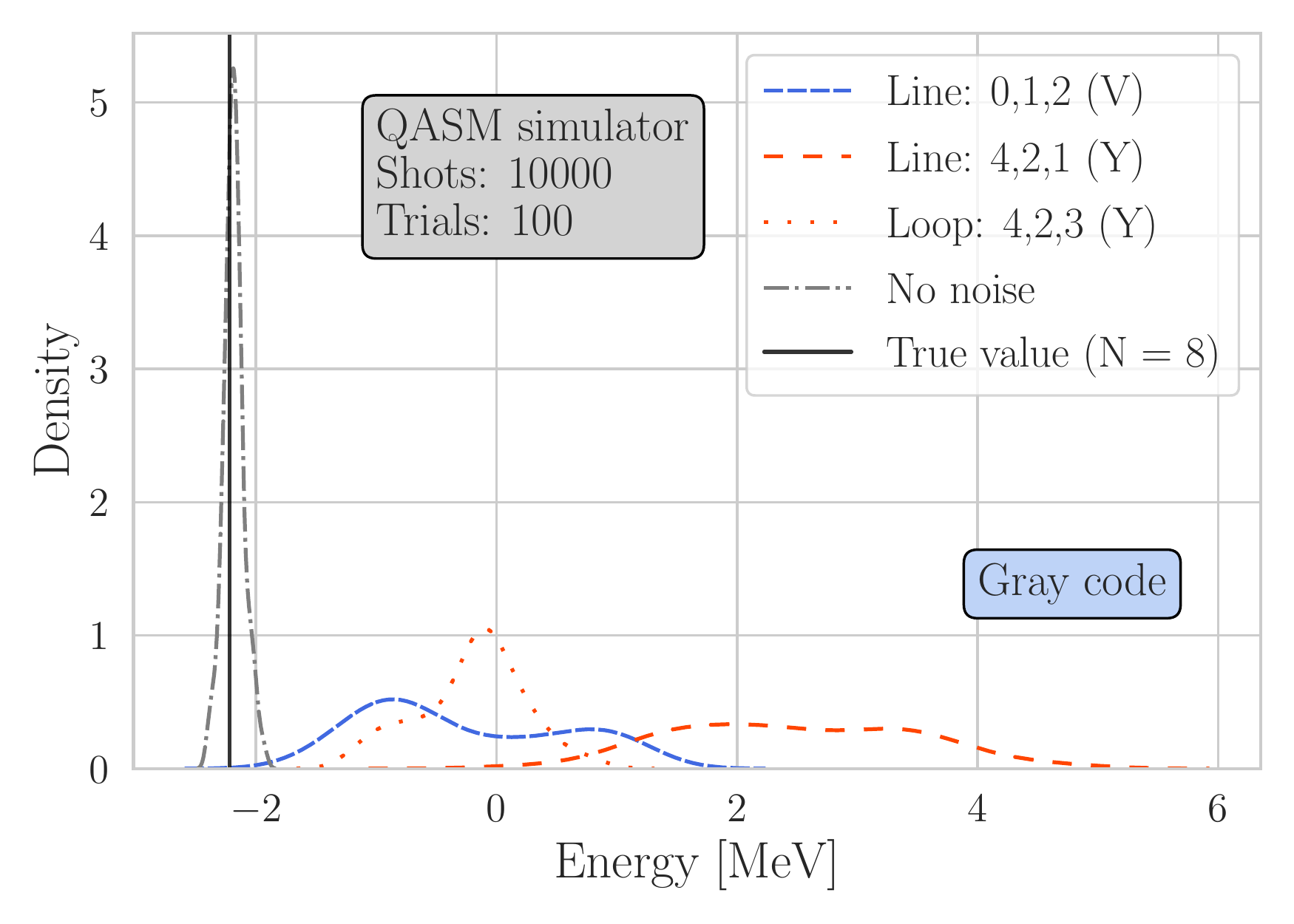}  
    \end{tabular}
    \caption{The distribution of VQE energies for QASM simulations of the deuteron Hamiltonian for $N=8$ with simulated qubits on the Vigo (V) and Yorktown (Y) devices using the Gray code encoding. The noisy results have measurement error mitigation applied. The legend specifies the layout of qubits on the simulated device.
    }
    \label{fig:yorktown_topologies}
\end{figure}

In addition to noise, the connectivity of current quantum computer architectures is a limiting factor in obtaining accurate results; i.e., the topology of the hardware graph can have a significant effect.
For example, when running a 3-qubit circuit (i.e., Figure \ref{fig:ansatz-8state}), on the IBM Q Yorktown device, the virtual qubits can be mapped to a line (with connections between \{0,1\} and \{1,2\}) or a loop (connecting \{0,2\} in addition).
Executing the Gray code circuit for 8 states requires a CNOT gate on \{0,2\}.
A line topology requires additional SWAP gates (each consisting of three CNOTs) and so experiences more noise.
A simulated demonstration of this effect is shown in Figure \ref{fig:yorktown_topologies}.
The loop topology is significantly better than the line topology on Yorktown.
The Vigo device cannot support a loop topology but due to its lower error rates a line topology is comparable to the loop topology on Yorktown.

\bibliography{main}


\end{document}